# Mechanical Design of Superconducting Accelerator Magnets

*F. Toral*[1]
CIEMAT, Madrid, Spain

**Abstract**
This paper is about the mechanical design of superconducting accelerator magnets. First, we give a brief review of the basic concepts and terms. In the following sections, we describe the particularities of the mechanical design of different types of superconducting accelerator magnets: solenoids, cos-theta, superferric, and toroids. Special attention is given to the pre-stress principle, which aims to avoid the appearance of tensile stresses in the superconducting coils. A case study on a compact superconducting cyclotron summarizes the main steps and the guidelines that should be followed for a proper mechanical design. Finally, we present some remarks on the measurement techniques.

## 1  Introduction

The designer of a superconducting magnet will be concerned about achieving a very good magnetic field quality and protecting the magnet in case of quench, but he or she should not forget that mechanical failures are the cause of performance loss in superconducting magnets, compared with that predicted by the electromagnetic computations.

Superconducting accelerator magnets are characterized by large fields and current densities. As a result, coils experience large stresses, which have three important effects.

i) Quench triggering: the most likely origin of quench is the release of stored elastic energy when part of the coil moves or a crack suddenly appears in the resin. Due to the low heat capacity of materials at low temperatures, the resulting energy deposition is able to increase the temperature of the superconductor above its critical value.

ii) Mechanical degradation of the coil or the support structure: if the applied forces/pressures are above a given threshold (yield strength), plastic deformation of the materials takes place.

iii) Field quality: the winding deformation may affect the field quality.

The parts of the magnet are produced and assembled at room temperature, but their working temperature is about −270ºC. The designer must consider carefully the differential thermal contractions of materials during cool-down and operation.

Taking into account the aforementioned aspects, the mechanical design will aim to:

i)  avoid tensile stresses on the superconductor;

ii) avoid mechanical degradation of the materials;

iii) study the magnet life cycle: assembly, cool-down, energizing, and quench.

---

[1] fernando.toral@ciemat.es

Figure 1 shows the strategy that the designer should follow during the mechanical analysis of a superconducting accelerator magnet.

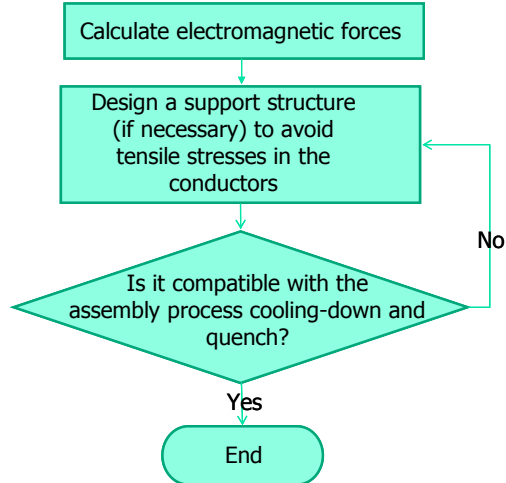

**Fig. 1:** Strategy for the mechanical design of superconducting magnets

## 2    Basic concepts

Some basic concepts from electromagnetism and elasticity theory will be reviewed in this section, with special attention paid to the particular expressions used in the following sections.

### 2.1    Electromagnetic forces

A charged particle $q$ moving with speed $v$ in the presence of an electric field $E$ and a magnetic field $B$ experiences the Lorentz force, which is given by

$$\vec{F}\,[\text{N}] = q(\vec{E} + \vec{v} \times \vec{B})\,. \tag{1}$$

In the same way, a conductor element carrying current density $j$ in the presence of a magnetic field $B$ will experience the force density

$$\vec{f}_\text{L}\,[\text{N}\cdot\text{m}^{-3}] = \vec{j} \times \vec{B}\,. \tag{2}$$

The Lorentz force is a body force, i.e. it acts on all the parts of the conductor, as does the gravitational force. The total force on a given body can be computed by integration:

$$\vec{F}_\text{L}\,[\text{N}] = \iiint \vec{f}_\text{L}\,\text{d}v\,. \tag{3}$$

The magnetic energy density $u$ stored in a region without magnetic materials ($\mu_\text{r} = 1$) in the presence of a magnetic field $B$ is

$$u\,[\text{J}\cdot\text{m}^{-3}] = \frac{\vec{B}\cdot\vec{H}}{2} = \frac{B^2}{2\mu_0}\,. \tag{4}$$

The total energy $U$ can be obtained by integration over all the space, by integration over the coil volume, or by knowing the so-called self-inductance $L$ of the magnet:

$$U\,[\text{J}] = \iiint_\text{all} \frac{\vec{B}\cdot\vec{H}}{2}\,\text{d}v = \iiint_\text{coil} \vec{A}\cdot\vec{j}\,\text{d}v = \frac{1}{2}LI^2\,. \tag{5}$$

The stored energy density may be understood as a 'magnetic pressure', $p_m$ (see Eq. (6)). In a current loop, the magnetic field line density is higher inside: the field lines try to expand the loop, like a gas in a container. The magnetic pressure is given by

$$p_m \left[ N \cdot m^{-2} \right] = \frac{B^2}{2\mu_0}. \tag{6}$$

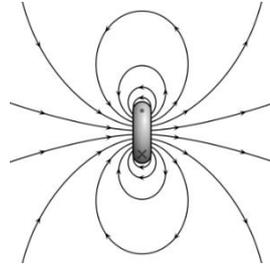

**Fig. 2:** Magnetic field lines created by a current loop (graph courtesy of www.answers.com)

## 2.2 Stress and strain

In continuum mechanics, stress is a physical quantity which expresses the internal pressure that neighbouring particles of a continuous material exert on each other. As shown in Fig. 3(a), when the forces are perpendicular to the plane, the stress is called normal stress ($\sigma$); when the forces are parallel to the plane, the stress is called shear stress ($\tau$). Stresses can be seen as the way a body resists the action (compression, tension, sliding) of an external force. A tensile (pulling) stress is considered as positive, and is associated with an elongation of the pulled body. As a consequence, a compressive (pushing) stress is negative and is associated with a body contraction [1]. The normal and shear stresses are given by

$$\sigma [\text{Pa}] = \frac{F_z}{A} \left[ N \cdot m^{-2} \right] \tag{7}$$

and

$$\tau [\text{Pa}] = \frac{F}{A} [N \cdot m]. \tag{8}$$

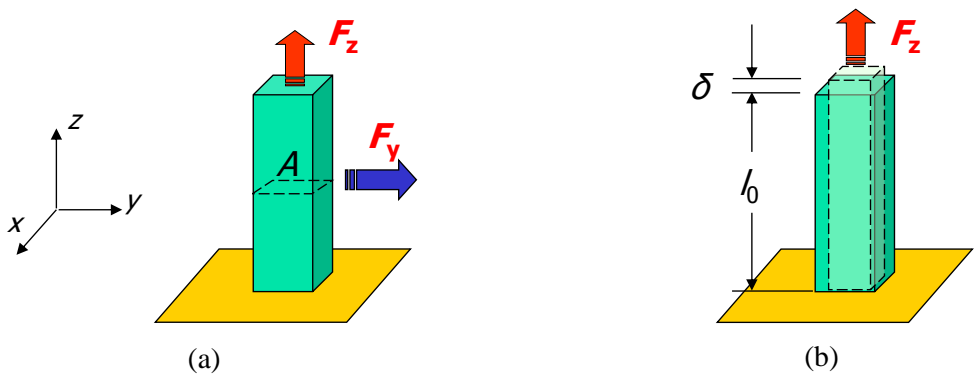

(a)          (b)
**Fig. 3:** (a) Normal and shear stresses. (b) Strain

A strain $\varepsilon$ is a normalized measurement of deformation representing the displacement $\delta$ between particles in the body relative to a reference length $l_0$ (see Fig. 3(b)):

$$\varepsilon = \frac{\delta}{l_0}. \tag{9}$$

According to *Hooke's law* (1678), within certain limits, the strain $\varepsilon$ of a bar is proportional to the exerted stress $\sigma$. The constant of proportionality is the elastic constant of the material, the so-called modulus of elasticity $E$, or Young modulus:

$$\varepsilon = \frac{\delta}{l_0} = \frac{\sigma}{E} = \frac{F}{AE}. \tag{10}$$

The *Poisson ratio* $\nu$ is the ratio of 'transverse' to 'axial' strain:

$$\nu = -\frac{\varepsilon_{\text{transversal}}}{\varepsilon_{\text{axial}}}. \tag{11}$$

When a body is compressed in one direction, it tends to elongate in the transverse direction. Conversely, when a body is elongated in one direction, it gets thinner in the other direction. The typical value is around 0.3.

A *shear modulus G* can be defined as the ratio of the shear stress $\tau$ and the shear strain $\gamma$:

$$G = \frac{\tau_{xy}}{\gamma_{xy}} = \frac{E}{2(1+\nu)}. \tag{12}$$

The proportionality between stress and strain is usually more complicated than suggested by Hooke's law (see Eq. (10)). Figure 4 shows the stress–strain graph for a typical material. The following individual points should be noted.

i) Point A shows the limit of proportionality. The first section of the curve is a straight line, in accordance with the linear behaviour described by Hooke's law.

ii) Point B is known as the yield point. It is usually defined as the point where the permanent deformation is 0.2%.

iii) Point C shows the ultimate strength. Beyond this point, the strain increases, even at lower stresses.

iv) Label D corresponds to the fracture point.

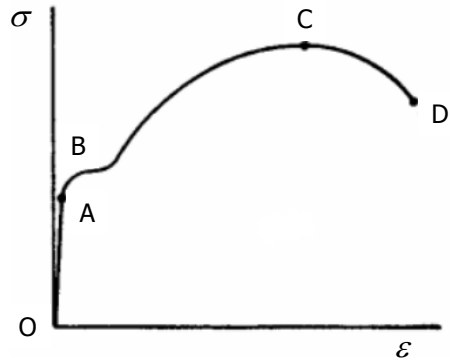

**Fig. 4:** Stress vs. strain graph for a typical material [1]

Several failure criteria are defined to estimate the failure/yield of structural components. One of the most broadly used is the equivalent (von Mises) stress $\sigma_v$, given by

$$\sigma_v = \sqrt{\frac{(\sigma_1 - \sigma_2)^2 + (\sigma_2 - \sigma_3)^2 + (\sigma_3 - \sigma_1)^2}{2}}, \tag{13}$$

where $\sigma_1$, $\sigma_2$, and $\sigma_3$ are the principal stresses.

## 2.3 Material properties

The properties of a material must be well known to perform a proper mechanical design. Superconducting windings are composites, i.e. mixtures of different materials. As a first approach, the magnet designer uses smeared-out properties of the winding, taking into account the volumetric fraction of each material and its distribution. For example, Table 1 shows the main properties of some materials commonly used in Nb−Ti coils. Figure 5(a) depicts a simple Nb−Ti coil wound with round wire, and Fig. 5(b) shows the model used to obtain the smeared-out properties (see Table 2), calculated by numerical methods. Material properties are strongly dependent on temperature. The designer may find some dispersion in the values depending on the source: Refs. [2], [3], and [4] can be used as general references.

**Table 1:** Physical properties of some typical materials used in Nb−Ti coils (at 4.2 K)

| Material | Young modulus (GPa) | Poisson ratio | Shear modulus (GPa) | Integrated contraction (296 to 4.2 K) | References |
|---|---|---|---|---|---|
| Nb–Ti | 77 | 0.3 | 20 | 1.87E-3 | [5] |
| Nb–Ti wire | 125 | 0.3 | 48 | 2.92E-3 | [6] |
| Copper | 138 | 0.34 | 52 | 3.15E-3 | [7] |
| Varnish insulation | 2.5 | 0.35 | 0.93 | 10.3E-3 | [8] |
| Epoxy | 7 | 0.28 | 2.75 | 6.40E-3 | [9] |

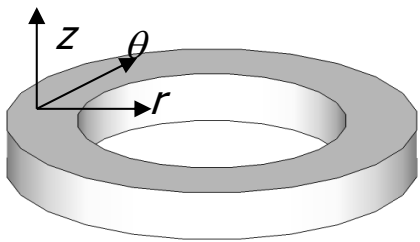
(a)

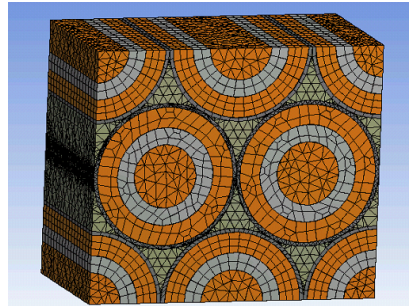
(b)

**Fig. 5**: (a) Simple solenoid winding. (b) Sub-model used to obtain the smeared-out mechanical properties

**Table 2:** Smeared-out mechanical properties of Nb−Ti winding (at 4.2 K)

| | Young modulus (GPa) | Poisson ratio | Shear modulus (GPa) | Integrated thermal contraction (296 to 4.2 K) |
|---|---|---|---|---|
| $\theta$ | 94 | | | 2.99E-3 |
| $r$ | 35 | | | 3.90E-3 |
| $z$ | 35 | | | 3.93E-3 |
| $r\theta$ | | 0.08 | | |
| $z\theta$ | | 0.08 | | |
| $rz$ | | 0.35 | 24 | |

Note the significant differences between the integrated thermal contractions of the materials shown in Table 1. The magnet assembly is always made at room temperature. Obviously, the magnet designer needs to analyze the induced stresses due to the different contraction coefficients of glued or clamped parts during the cooling down. Special attention must be paid to the degradation of insulating materials.

## 3 Solenoids

Solenoids will be the first type of coils to be reviewed, due to their simple geometry. In this case, the analytical expressions are easily deduced. Numerical methods will then be described, and some remarks on their advantages and risks will be included.

### 3.1 Thin-wall solenoids

In an infinitely long solenoid carrying current density $j$, the field inside is uniform and outside is zero. Lorentz forces push the coil outwards in a purely radial direction, creating a hoop stress $\sigma_\theta$ on the wires. We assume that the wall thickness is very small.

First, using Ampère's law, we can compute the field inside the solenoid:

$$\oint \vec{B} \cdot d\vec{l} = \mu_0 I \Rightarrow B_0 \Delta z = \mu_0 j w \Delta z . \tag{14}$$

Assuming that the coil's average field is $B_0/2$, the magnetic pressure $p_\mathrm{m}$ is given by the distributed Lorentz force $F_\mathrm{L}$:

$$F_{\mathrm{L}r}\vec{u}_r = f_{\mathrm{L}r}\left(a\Delta\theta\Delta zw\right)\vec{u}_r = j\frac{B_0}{2}\left(a\Delta\theta\Delta zw\right)\vec{u}_r = p_\mathrm{m}\left(a\Delta\theta\Delta z\right)\vec{u}_r , \tag{15}$$

$$p_\mathrm{m} = \frac{B_0^2}{2\mu_0} , \tag{16}$$

where $f_{\mathrm{L}r}$ is the density of the Lorentz force in the radial direction. It is important to note that *the magnetic pressure increases with the square of the field*. For example, in an infinitely long solenoid with a central field of 10 T, the windings undergo a pressure of 398 atm!

The simplest stress calculation is based on the assumption that each turn acts independently of its neighbours. Based on the equilibrium of forces on half a solenoid, as shown in Fig. 6(b), one may compute the hoop stress $\sigma_\theta$:

$$\begin{aligned} 2F &= \int_{-\pi/2}^{\pi/2} f_{\mathrm{L}r} \cos\theta\, aw\, d\theta = 2 f_{\mathrm{L}r} aw, \\ F &= f_{\mathrm{L}r} aw = p_\mathrm{m} a \Rightarrow \sigma_\theta w = \frac{B_0^2}{2\mu_0} a \Rightarrow \sigma_\theta = \frac{B_0^2}{2\mu_0}\frac{a}{w} = p_\mathrm{m}\frac{a}{w}. \end{aligned} \tag{17}$$

In general, the peak stress occurs in the innermost turn, where the magnetic field is also maximum. Using Eq. (2), and assuming that $B$ is the field at the innermost turn, located at radius $a$, the peak stress can be calculated as

$$\sigma_{\max} = BJa \propto J^2 . \tag{18}$$

The reader should note that *the peak stress increases with the square of the current density*. In our example, assuming an inner radius of 10 cm and a thickness of 10 mm, the peak stress is about 400 MPa. It is too high for a Nb$_3$Sn winding (yield stress ~150 MPa) and possibly even for a Nb–Ti winding (yield stress ~500 MPa), assuming a filling factor of 70%. In that case, how should one build a robust solenoid able to create a central field of 10 T? The solution is to exert a pre-stress on the winding that correspondingly decreases the tensile azimuthal stress $\sigma_\theta$.

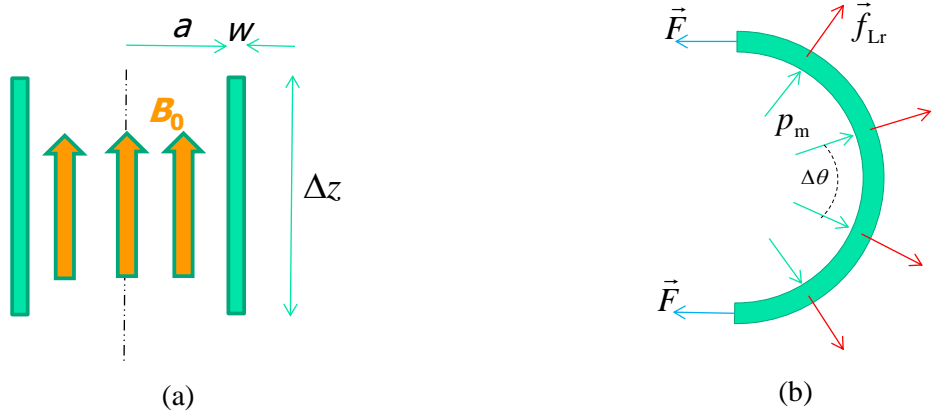

(a)  (b)

**Fig. 6:** (a) Uniform magnetic field inside a solenoid. (b) Radial pressure on the solenoid due to Lorentz forces

### 3.2 Thick-wall solenoids

Figure 7(a) shows the magnetic field map of a typical thick solenoid winding, and Fig. 7(b) depicts the Lorentz forces when it is energized. The wall thickness is not negligible compared with the length. The electromagnetic forces tend to push the coil:

– outwards in the radial direction ($F_r > 0$);

– towards the mid-plane in the axial direction ($F_y < 0$ in the upper half coil, and the opposite in the lower half).

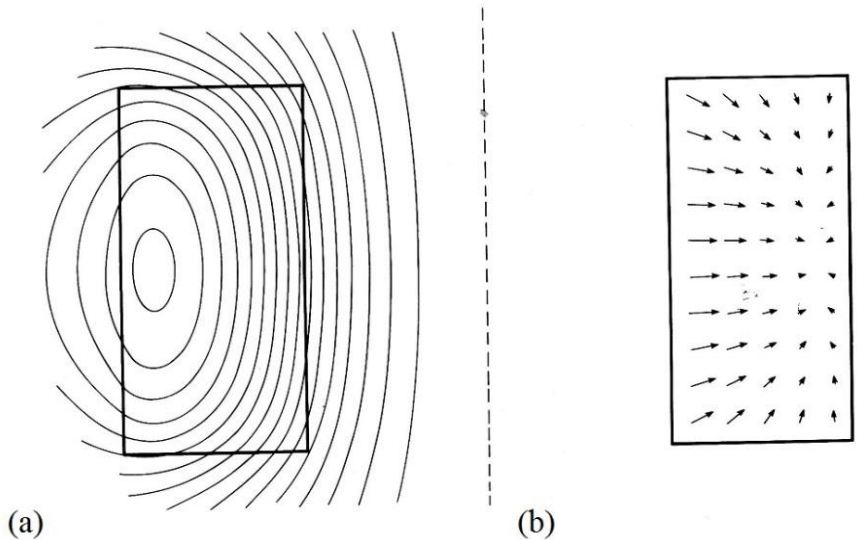

(a)  (b)

**Fig. 7:** (a) Magnetic field lines created by a thick solenoid winding. (b) Lorentz forces (graphs from [10])

Figure 8 shows the azimuthal stress distribution for long solenoids with two different shape factors. The shape factor is the ratio of the outer and the inner radii. The label $\sigma'_\theta$ is given to the curves depicting the hoop stress when we assume that the turns act independently, which is a poor approximation. The label $\sigma_\theta$ shows the hoop stress calculated when we assume that adjacent turns press on each other, developing radial stresses. Note that thin solenoids perform negative radial stresses, whereas thick solenoids show regions with positive radial stresses, i.e. tensile stresses. In the latter case, there is a risk of resin cracking or wire movement, which could trigger a quench. In summary, long and thin solenoids are mechanically more stable than thick ones.

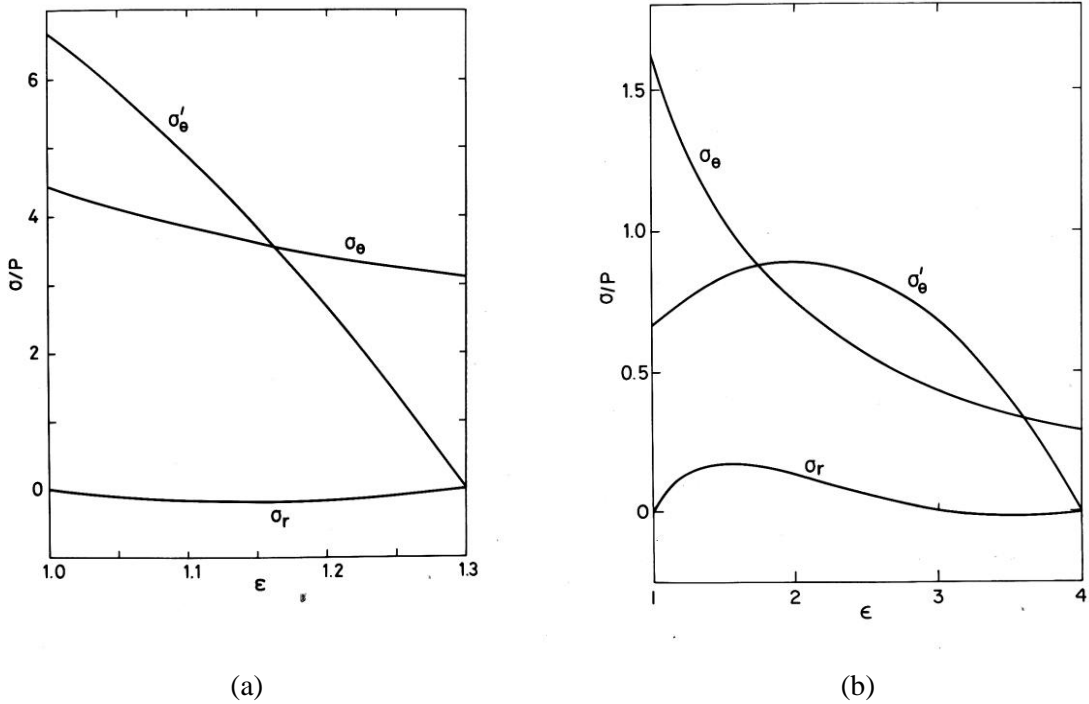

(a)                                  (b)

**Fig. 8:** (a) Azimuthal stress $\sigma_\theta$ distribution in long solenoids with shape factors (a) 1.3, (b) 4.0 [10]

### 3.3 Application example

Figure 9(b) shows the set-up prepared to test a superconducting solenoid with a superconducting switch. This magnet should be used as a mock-up of the Alpha Magnetic Spectrometer (AMS) main magnet to test its power supply [11]. It is wound with an enamelled Nb-Ti round wire. In the first cool-down, the learning curve was very slow (see Fig. 10). The magnet quenched at very low currents, below half the short sample limit, while the nominal working point was at 75% on the load line. After warm-up, the (re)-training did not improve; indeed, a slight de-training (quench current lower than in the previous quenches) was observed in the first quench. It is clear that a mechanical problem limits the magnet performance. It was decided to stop the training tests and to analyze the mechanics carefully.

| | |
|---|---|
| Length | 123.75 mm |
| Inner diameter | 188.72 mm |
| Outer diameter | 215 mm |
| Number of turns | 1782 |
| Nominal current | 450 A |
| Peak field | 5.89 T |
| Bore field | 4.25 T |
| Current density | 493 A·mm$^{-2}$ |
| Working point | 75% |
| Self inductance | 0.5 H |

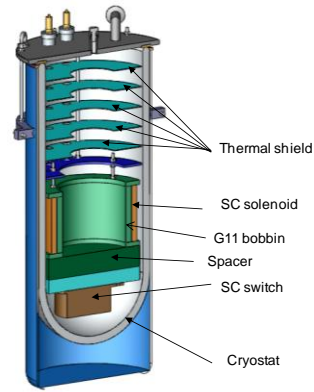

(a)        (b)

**Fig. 9:** (a) Solenoid parameters. (b) Solenoid test set-up

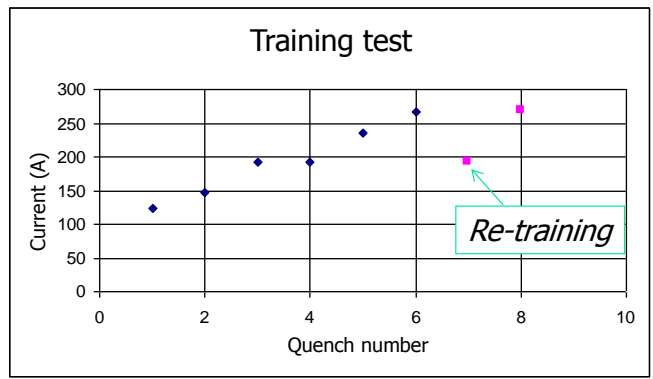

**Fig. 10:** Learning curve of the superconducting solenoid: first training and re-training

The solenoid was wound on a G10 bobbin and wet impregnated with epoxy resin. The integrated (from 300 down to 4.2 K) thermal contraction coefficient of the glass fibre and the winding are quite different, 280E-5 and 392E-5, respectively. The detailed mechanical analysis was not made before the magnet fabrication because it was not considered necessary for this small test coil. However, even for such a small magnet, the wrong mechanical design may spoil the performance. When the Finite Element Method (FEM) numerical analysis was performed, tensile stresses up to 46 MPa were detected in the coil ends (see Fig. 11(a)). These are able to crack the epoxy resin, triggering the premature quenches. It was decided to turn the bobbin core, and to split it into two different parts (see Fig. 11(b)). Now the coil is working mainly under compression. When the magnet was cooled down, the training improved significantly (see Fig. 12, yellow and light blue dots). It finally reached the nominal current after a few quenches.

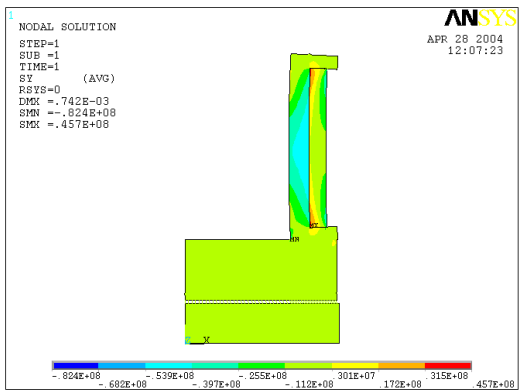
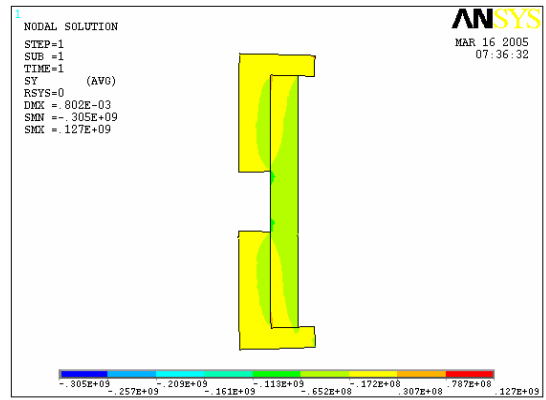

(a)                   (b)

**Fig. 11:** Axial (vertical) stress: (a) continuous G10 bobbin; (b) split G10 bobbin

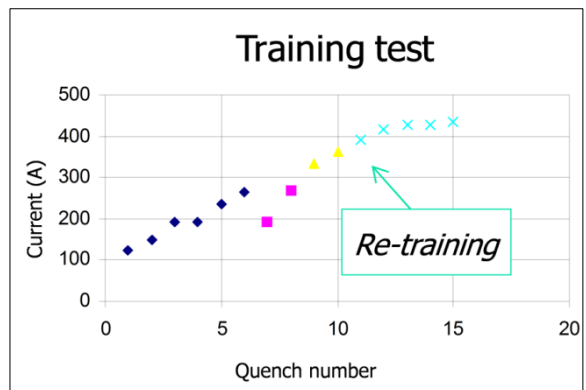

**Fig. 12:** Learning curve of the superconducting solenoid: complete training

### 3.4 Numerical vs. analytical methods

In this section, analytical and numerical methods are applied to the calculation of the stresses present in solenoid windings. Analytical expressions are valid only for simplified models, while numerical methods are able to model very precisely the real magnet, i.e. the

- anisotropic material properties;
- complicated/detailed geometry;
- 'sophisticated' boundary conditions: sliding/contact surfaces, joints;
- load steps: assembly, cooling down, energizing;
- transient problems.

There is a common temptation to forget about the analytical approach and start the analysis directly with the numerical simulations. This is not the most effective strategy, however. The analytical methods have to be used first, because they allow us to:

- understand the problem and the physics behind it;
- make a first estimate of the solution;
- simplify the numerical simulation;
- check and *understand* the results of the numerical simulations.

## 4 Cos-theta accelerator magnets

This type of magnet is the most common in particle accelerators, since the superconductor efficiency is very high (the current distribution is very close to the aperture) and permits very high magnetic fields to be achieved. The geometry is relatively complicated, especially at the coil ends.

### 4.1 Lorentz forces

The Lorentz forces in an *n*-pole magnet tend to push the coil:

- towards the mid plane in the vertical/azimuthal direction ($F_y$, $F_\theta < 0$);
- outwards in the radial–horizontal direction ($F_x$, $F_r > 0$).

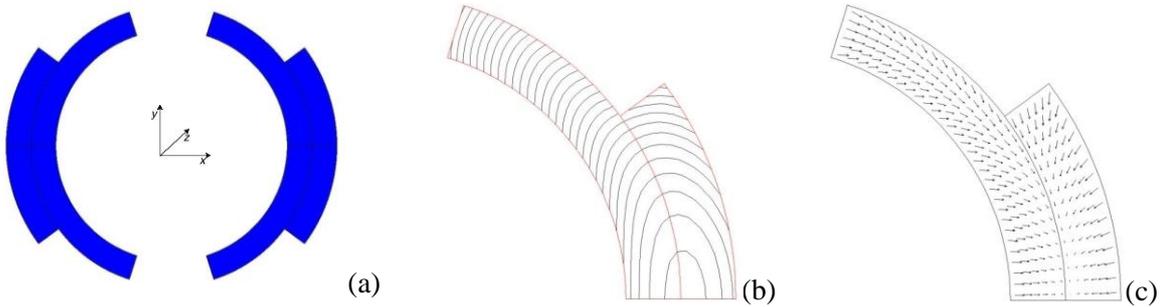

                (a)                 (b)                 (c)

**Fig. 13:** (a) Cos-theta dipole winding; (b) field map on the coil (c) electromagnetic forces [12]

At the coil ends, the Lorentz forces tend to make the coil longer. The forces are pointing outwards in the longitudinal direction ($F_z > 0$).

In short, the electromagnetic forces try to expand the coil, as in a current loop. However, the coil by itself is unable to support the magnetic forces in tension. These forces must be counteracted by an external support structure.

In order to estimate the value of these forces, three different approximations can be considered for any *n*-pole magnet (see Fig. 14) [12]:

- *Thin shell*: the current density may be expressed as $J = J_0 \cos n\theta$ (ampere per unit circumference), where $J_0$ is a constant. This is the simplest model. It allows us to estimate orders of magnitude and proportionalities.

- *Thick shell*: the current density may be written as $J = J_0 \cos n\theta$ (ampere per unit area). This model may be used to get a first-order estimate of forces and stresses.

- *Sector*: the current density $J$ is constant (ampere per unit area). The sector spans an angle $\theta = 60º(30º)$ for a dipole(quadrupole), to make zero the first magnetic field harmonic. This model may be used to obtain a first-order estimate of forces and stresses.

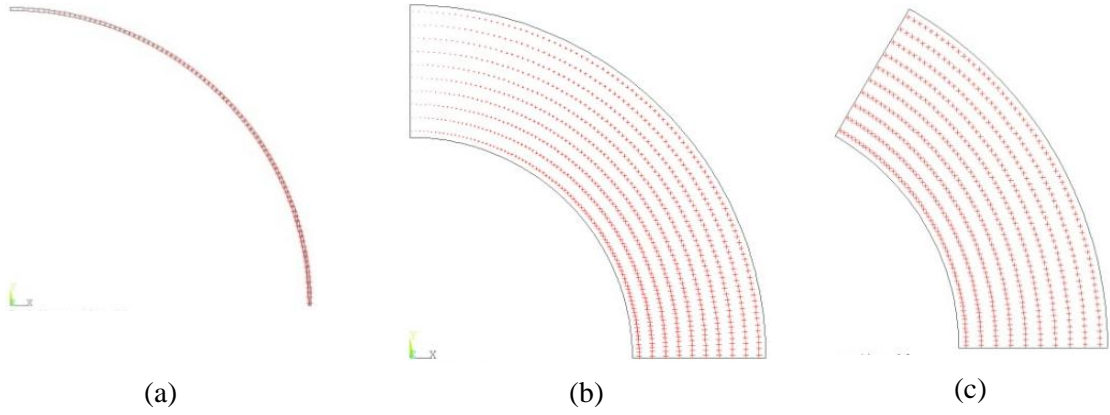

(a)            (b)            (c)

**Fig. 14:** Current density in the different winding approximations: (a) thin shell, (b) thick shell, and (c) sector

### 4.1.1 *Electromagnetic forces on a thin shell*

Beth's theorem states that the complex force on a current element (per unit length in the longitudinal direction $z$) is equal to the line integral of magnetic pressure around the boundary of that element in the complex plane:

$$\vec{F} = F_y + iF_x = -\oint \frac{B^2}{2\mu_0} dz . \tag{19}$$

For a cylindrical current sheet, the total force [N·m$^{-1}$] on half a coil is [10]:

$$\vec{F} = \frac{1}{2\mu_0} \int_0^{\pi/2n} \left( B_{in}^2 - B_{out}^2 \right) i a e^{i\theta} \, d\theta . \tag{20}$$

If the current density is given by $J = J_0 \cos n\theta$ [A·m$^{-1}$], and assuming an iron with infinite permeability $\mu = \infty$ placed at radius $R$ (see Fig. 15), the density force $f$ [N·m$^{-2}$] is given by [12–14]

$$\vec{f} = f_x + if_y = \frac{\mu_0 J_0^2}{8} \left\{ \left[ 1 + 2\left(\frac{a}{R}\right)^{2n} \right] e^{-i\theta(2n-1)} - e^{i\theta(2n+1)} + 2\left(\frac{a}{R}\right)^{2n} e^{i\theta} \right\} . \tag{21}$$

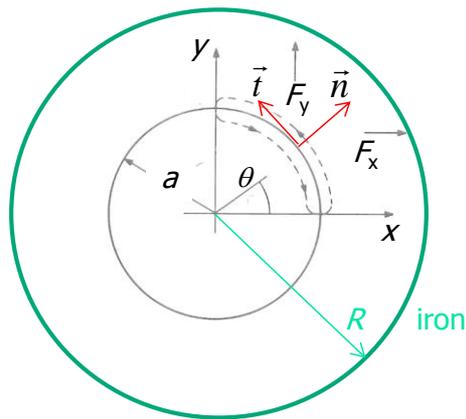

**Fig. 15:** Electromagnetic forces on a cylindrical current sheet surrounded by iron with infinite permeability

The tangential and normal components of the density force may be obtained by calculating the dot products with the tangent and normal unit vectors, $t$ and $n$, respectively:

$$f_r = \vec{f} \cdot \vec{n} = \vec{f} \cdot e^{i\theta} = \frac{\mu_0 J_0^2}{4} \left(\frac{a}{R}\right)^{2n} \left[ 1 + \cos(2n\theta) \right], \tag{22}$$

$$f_\theta = \vec{f} \cdot \vec{t} = \vec{f} \cdot e^{i(\pi/2+\theta)} = -\frac{\mu_0 J_0^2}{4}\left[1+\left(\frac{a}{R}\right)^{2n}\right]\sin(2n\theta). \qquad (23)$$

For a dipole, the force on half the coil is given by

$$F_x\left[\text{N}\cdot\text{m}^{-1}\right] = \int_0^{\pi/2}(f_r\cos\theta - f_\theta\sin\theta)a\,d\theta = \frac{\mu_0 J_0^2}{2}\left[\frac{1}{3}+\left(\frac{a}{R}\right)^2\right]a\,; \qquad (24)$$

$$F_y\left[\text{N}\cdot\text{m}^{-1}\right] = \int_0^{\pi/2}(f_r\sin\theta + f_\theta\cos\theta)a\,d\theta = -\frac{\mu_0 J_0^2}{2}\frac{1}{3}a\,. \qquad (25)$$

*It is proportional to the bore radius and the square of the current density (and field).* The term containing the iron radius $R$ is the contribution from the iron, which can be easily distinguished from the contribution from the conduction current.

In a rigid magnet structure, the force determines an azimuthal displacement of the coil and creates a separation at the pole (see Fig. 26). The structure should withstand $F_x$. Meanwhile, $F_y$ provides a compression on the coil itself, with a maximum stress at the mid-plane. If one thinks of a coil working as a "roman arch", where all the hoop forces $f_\theta$ accumulate on the mid-plane, the total force $F_\theta$ transmitted on the mid-plane per unit length of the magnet is

$$F_\theta = \int_0^{\pi/2} f_\theta a\,d\theta = -\frac{B_y^2}{\mu_0}a\,. \qquad (26)$$

Furthermore, one can consider a real winding as a set of current sheets and solve the problem by superposition. This method allows us to compute the magnetic field, the stored magnetic energy, and the electromagnetic forces [13]. As an application example, Fig. 16 shows one coil of a corrector quadrupole prototype magnet developed for LHC, the so-called MQTL. It is split into a set of thin shells. Table 3 shows good agreement in the results using three different methods: the superposition of thin shells, a BEM–FEM numerical calculation (ROXIE [15]), and FEM numerical calculation.

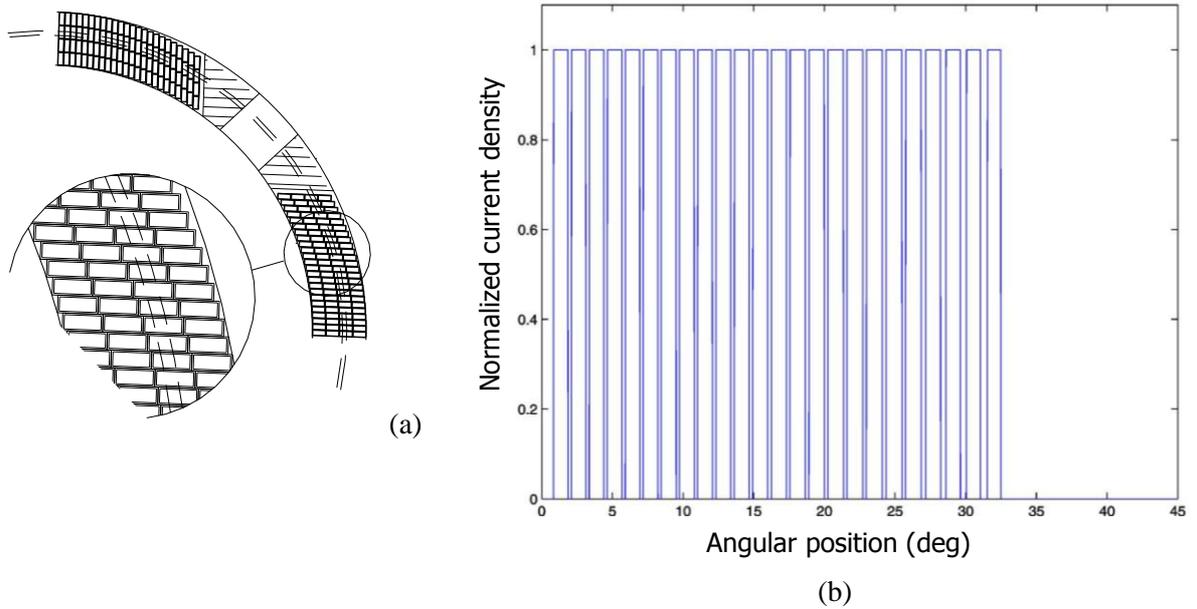

**Fig. 16:** (a) MQTL winding: detailed view of the subdivision in thin shells. (b) Current density in one of the thin shells.

Table 3: Magnetic field, stored energy, and forces in MQTL magnet using different computation methods

| Magnitude | Thin shells | ROXIE | FEM | Units |
|---|---|---|---|---|
| Gradient | 2.99 | 3.00 | 2.99 | T·m$^{-1}$ |
| $b_6$ | 98.4854 | 98.4616 | 98.1640 | 1E-4 |
| $b_{10}$ | 1.1899 | 1.1871 | 1.4283 | 1E-4 |
| $b_{14}$ | 0.0152 | 0.0152 | 0.2352 | 1E-4 |
| $B_{max}$ | 0.366 | 0.368 | 0.379 | T |
| $L$ | 0.0547 | 0.0546 | 0.0548 | mH·m$^{-1}$ |
| $F_x$ | 51.74 | 50.30 | 48.955 | N |
| $F_y$ | −118.26 | −116.57 | −115.27 | N |

### 4.1.2  Electromagnetic forces on a thick shell

Assuming that the current density is $J = J_0 \cos n\theta$, where $J_0$ is measured in A·m$^{-2}$, the shell inner radius is $a_1$, the outer radius is $a_2$, and no iron is present, the radial and azimuthal components of the magnetic field $B_i$ inside the aperture of an $n$-pole magnet are given by [12, 14]

$$B_{ri} = -\frac{\mu_0 J_0}{2} r^{n-1} \left( \frac{a_2^{2-n} - a_1^{2-n}}{2-n} \right) \sin n\theta, \tag{27}$$

$$B_{\theta i} = -\frac{\mu_0 J_0}{2} r^{n-1} \left( \frac{a_2^{2-n} - a_1^{2-n}}{2-n} \right) \cos n\theta. \tag{28}$$

The radial and azimuthal components of the field at the coil may be written as follows:

$$B_r = -\frac{\mu_0 J_0}{2} \left[ r^{n-1} \left( \frac{a_2^{2-n} - r^{2-n}}{2-n} \right) + \frac{1}{2+n} \left( \frac{r^{2+n} - a_1^{2+n}}{r^{1+n}} \right) \right] \sin n\theta, \tag{29}$$

$$B_\theta = -\frac{\mu_0 J_0}{2} \left[ r^{n-1} \left( \frac{a_2^{2-n} - r^{2-n}}{2-n} \right) - \frac{1}{2+n} \left( \frac{r^{2+n} - a_1^{2+n}}{r^{1+n}} \right) \right] \cos n\theta. \tag{30}$$

The radial and azimuthal components of the electromagnetic force density (measured in N·m$^{-3}$) acting on the coil are:

$$f_r = -B_\theta J = \frac{\mu_0 J_0^2}{2} \left[ r^{n-1} \left( \frac{a_2^{2-n} - r^{2-n}}{2-n} \right) - \frac{1}{2+n} \left( \frac{r^{2+n} - a_1^{2+n}}{r^{1+n}} \right) \right] \cos^2 n\theta, \tag{31}$$

$$f_\theta = B_r J = -\frac{\mu_0 J_0^2}{2} \left[ r^{n-1} \left( \frac{a_2^{2-n} - r^{2-n}}{2-n} \right) + \frac{1}{2+n} \left( \frac{r^{2+n} - a_1^{2+n}}{r^{1+n}} \right) \right] \sin n\theta \cos n\theta. \tag{32}$$

The Cartesian components of the Lorentz force density may be computed using the following expressions:

$$f_x = f_r \cos\theta - f_\theta \sin\theta, \tag{33}$$

$$f_y = f_r \sin\theta + f_\theta \cos\theta. \tag{34}$$

In the particular case of a dipole, the field inside the coil is given by

$$B_y = -\frac{\mu_0 J_0}{2}(a_2 - a_1). \tag{35}$$

The components of the total force acting on the coil per unit length are given by

$$F_x = \frac{\mu_0 J_0^2}{2}\left[\frac{7}{54}a_2^3 + \frac{1}{9}\left(\ln\frac{a_2}{a_1} + \frac{10}{3}\right)a_1^3 - \frac{1}{2}a_2 a_1^2\right], \tag{36}$$

$$F_y = -\frac{\mu_0 J_0^2}{2}\left[\frac{2}{27}a_2^3 + \frac{2}{9}\left(\ln\frac{a_1}{a_2} - \frac{1}{3}\right)a_1^3\right]. \tag{37}$$

A very simple approximation of the maximum stress at the mid-plane is given by

$$\sigma_y = \frac{F_y}{b-a}. \tag{38}$$

### 4.1.3 *Electromagnetic forces on a sector coil*

Assuming a uniform current density $J = J_0$ perpendicular to the cross-section plane, inner radius $a_1$, outer radius $a_2$, a span angle $\phi$ such that the first allowed field harmonic is null (i.e. $\phi = 60°$ for a dipole), and no iron is present, the polar components of the magnetic field inside the aperture are [12, 14] as follows:

$$B_{ri} = -\frac{2\mu_0 J_0}{\pi}\left[(a_2 - a_1)\sin\phi\sin\theta + \sum_{n=1}^{\infty}\frac{r^{2n}}{(2n+1)(2n-1)}\left(\frac{1}{a_1^{n-1}} - \frac{1}{a_2^{n-1}}\right)\sin(2n+1)\phi\sin(2n+1)\theta\right], \tag{39}$$

$$B_{\theta i} = -\frac{2\mu_0 J_0}{\pi}\left[(a_2 - a_1)\sin\phi\cos\theta + \sum_{n=1}^{\infty}\frac{r^{2n}}{(2n+1)(2n-1)}\left(\frac{1}{a_1^{n-1}} - \frac{1}{a_2^{n-1}}\right)\sin(2n+1)\phi\cos(2n+1)\theta\right]. \tag{40}$$

The radial and azimuthal components of the field in the coil are given by

$$B_r = -\frac{2\mu_0 J_0}{\pi}\left\{(a_2 - r)\sin\phi\sin\theta + \sum_{n=1}^{\infty}\left[1 - \left(\frac{a_1}{r}\right)^{2n+1}\right]\frac{r}{(2n+1)(2n-1)}\sin(2n-1)\phi\sin(2n-1)\theta\right\}, \tag{41}$$

$$B_\theta = -\frac{2\mu_0 J_0}{\pi}\left\{(a_2 - r)\sin\phi\cos\theta - \sum_{n=1}^{\infty}\left[1 - \left(\frac{a_1}{r}\right)^{2n+1}\right]\frac{r}{(2n+1)(2n-1)}\sin(2n-1)\phi\cos(2n-1)\theta\right\}. \tag{42}$$

In the case of a dipole, the polar components of the Lorentz force density are given by

$$f_r = -B_\theta J = +\frac{2\mu_0 J_0^2}{\pi}\sin\phi\left[(a_2 - r) - \frac{r^3 - a_1^3}{3r^2}\right]\cos\theta, \tag{43}$$

$$f_\theta = B_r J = -\frac{2\mu_0 J_0^2}{\pi}\sin\phi\left[(a_2 - r) + \frac{r^3 - a_1^3}{3r^2}\right]\sin\theta. \tag{44}$$

The Cartesian components of the total force acting on the coil per unit length are given by

$$F_x = +\frac{2\mu_0 J_0^2}{\pi}\frac{\sqrt{3}}{2}\left[\frac{2\pi-\sqrt{3}}{36}a_2^3 + \frac{\sqrt{3}}{12}\ln\frac{a_2}{a_1}a_1^3 + \frac{4\pi+\sqrt{3}}{36}a_1^3 - \frac{\pi}{6}a_2 a_1^2\right], \qquad (45)$$

$$F_y = -\frac{2\mu_0 J_0^2}{\pi}\frac{\sqrt{3}}{2}\left[\frac{1}{12}a_2^3 + \frac{1}{4}\ln\frac{a_1}{a_2}a_1^3 - \frac{1}{12}a_1^3\right]. \qquad (46)$$

### *4.1.4 Axial electromagnetic forces on the coil ends*

The virtual displacement principle establishes that the variation of stored magnetic energy $U$ with the magnet length equals the axial force $F_z$ pulling from the ends, as long as the rest of the dimensions are kept constant:

$$F_z = \frac{\partial U}{\partial z}. \qquad (47)$$

That is, the stored magnetic energy per unit length equals the axial force: in the LHC main dipoles, it is about 125 kN per coil end.

If the coil is approximated as a thin shell, the axial force $F_z$ may be written as follows:

$$F_z = \frac{\mu_0 \pi}{4n}\left[1+\left(\frac{a}{R}\right)^{2n}\right]J_0^2 a^2 = \frac{B_y^2}{\mu_0}\pi a^2. \qquad (48)$$

*The axial force in a dipole increases with the square of the magnetic field and the aperture.* For the same current density, the end forces on a quadrupole coil are half those measured in a dipole. Similar expressions for thick shell and sector approximations may be found in Ref. [12].

## 4.2 Pre-stress

As pointed out in the previous sections, one of the main concerns of the mechanical designer is to avoid tensile stresses on the superconducting conductors when they are powered. The classical solution is to apply a pre-compression. This method was implemented in ancient times, for example in the Roman arch bridge (Fig. 17(a)). In the case of cos-theta winding configurations, the external structure usually applies a radial inward compression (Fig. 17(b)), which is transformed into an azimuthal compression inside the coil which counteracts the formation of tensile stresses that would otherwise appear under the action of the electromagnetic forces.

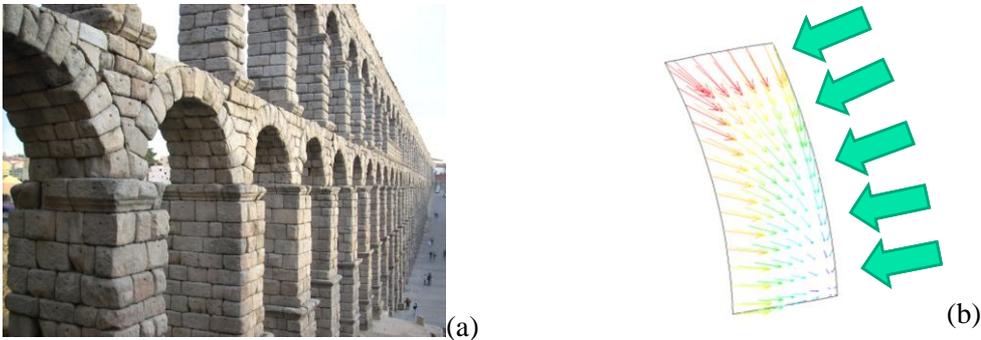

**Fig. 17:** (a) Roman arch aqueduct in Segovia (Spain) (courtesy of http://commons.wikimedia.org). (b) Electromagnetic forces on a quadrupole coil, counteracted by a radial inward compression.

The simplest structure that will provide external pre-compression is a cylindrical shell. It is usually made from aluminium, since its high thermal contraction eases the assembly (less interference is necessary to provide a given pressure at cold conditions), as will be seen later. The maximum stress in an aluminium shell at cool-down is about 200–300 MPa. As a first approach, one can assume that the radial Lorentz force behaves as a uniform pressure (see Fig. 18(a)) or, alternatively, take its horizontal component:

$$\sigma_\theta [\text{MPa}] = \frac{F[\text{N} \cdot \text{m}]}{\delta} = \frac{P \cdot a}{\delta}, \qquad \sigma_\theta [\text{MPa}] = \frac{F_{Lx}}{\delta}. \qquad (49)$$

For *n*-pole magnets, one can compute the bending moment in a thin cylinder under radial forces separated by an angle of $2\theta$ and the corresponding hoop stress as [16]:

$$M = \frac{Fa}{2}\left(\frac{\cos(x)}{\sin(\theta)} - \frac{1-\delta/a}{\theta}\right),$$

$$\sigma_\theta = \frac{6M}{\delta^2}. \qquad (50)$$

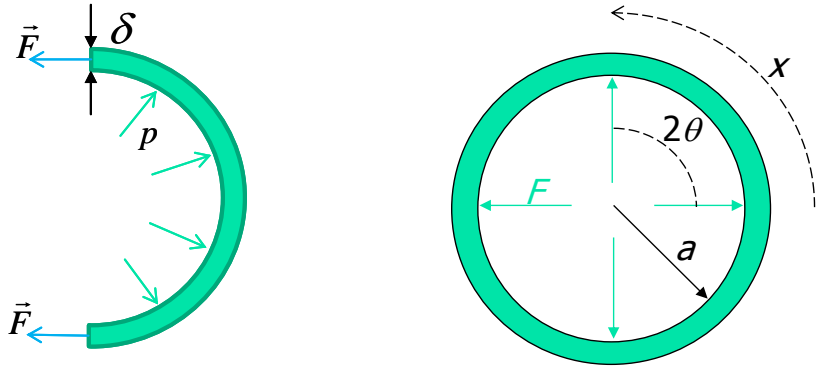

**Fig. 18:** (a) Cylindrical shell under outward radial pressure. (b) Cylindrical shell for an *n*-pole magnet

The maximum compressive stress at the coil must be checked. It usually takes place at the mid-plane, as shown in previous sections. If it is too high for the insulation, the most common solution is to reduce the current density accordingly.

In the following sections, some particular aspects of pre-stress application in real magnets will be reviewed depending on the magnet field value.

### *4.2.1 Low field magnets*

We will refer to magnets as 'low field' magnets if the coil peak field is below 4 T. Coils are usually made with monolithic wires and then fully impregnated. This is, for example, the situation for most of the LHC corrector magnets. The easiest way to provide the pre-compression is by means of an outer aluminium shrinking cylinder. It is very convenient to place the iron as close as possible to the coils, i.e. inside the shell, to enhance the field. However, the iron cannot be constructed as a hollow cylinder or ring laminations, because the iron contracts less than the aluminium and the coils would become loose inside the iron yoke. A clever lamination layout, the so-called 'scissors' lamination, was developed at CERN [17]: eccentric paired laminations with different orientations apply the inward pressure alternatively on neighbouring coils (see Fig. 19). An additional advantage of this system is its low price for series production, as the laminations can be accurately produced by fine blanking.

The MQTL was the longest corrector magnet produced for LHC using scissor-type laminations. Some interesting lessons can be drawn from the prototyping phase [18]. First, it is worth noting that a few holes have been drilled in the iron to maintain a good field quality even with moderate iron saturation. The first allowed multipole, $b_6$, varies with the current when the iron becomes saturated. Holes in the iron help to achieve a similar magnetic field map at low and high operating currents, i.e.

the variation of $b_6$ with current is reduced. Figure 20 shows that elliptical holes are a better choice than circular ones from the point of view of the mechanics, since the concentration of radial stresses on the inner edge of the hole is lower.

The second interesting feature appeared during the training test of the second prototype (see Fig. 21). The learning curve was very slow (see curves with the label 'V1'), starting with a first quench at very low current, about 220 A, while nominal current was 550 A. Also, there was no improvement when the magnet was cooled down from 4 to 1.9 K: one can conclude that there is a mechanical problem which limits the magnet performance. The interference of the shrinking cylinder was increased (see curves with the label 'ModA') to provide a higher pre-stress to the coils, but the magnet behaviour did not improve significantly. The de-training that happened occasionally suggests 'slip-stick' movements between the iron laminations and the coil package due to axial electromagnetic forces as a likely origin of the poor training.

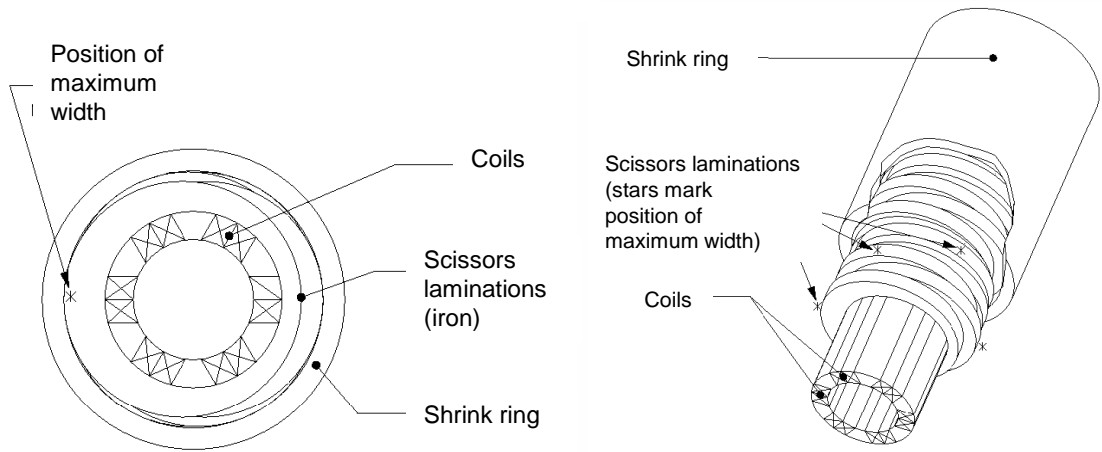

**Fig. 19:** Scissor laminations to provide pre-stress on the coils of LHC corrector magnets

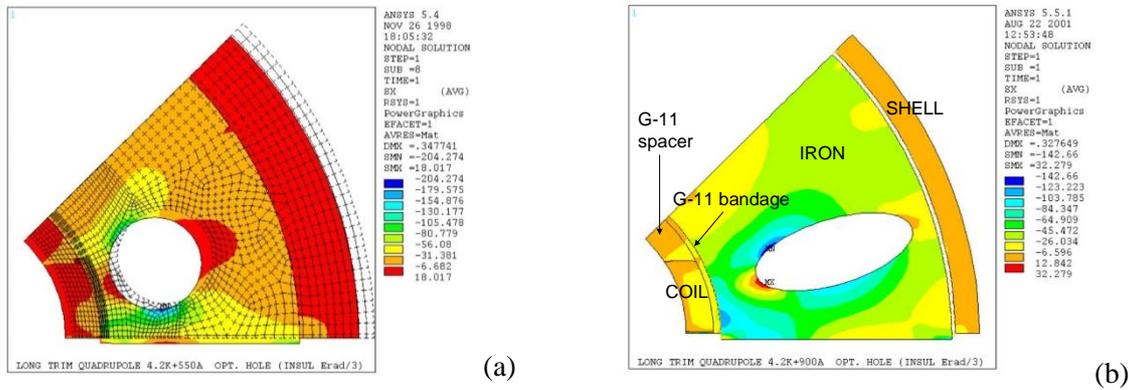

**Fig. 20:** Radial stress distribution for circular (a) and elliptical (b) iron holes

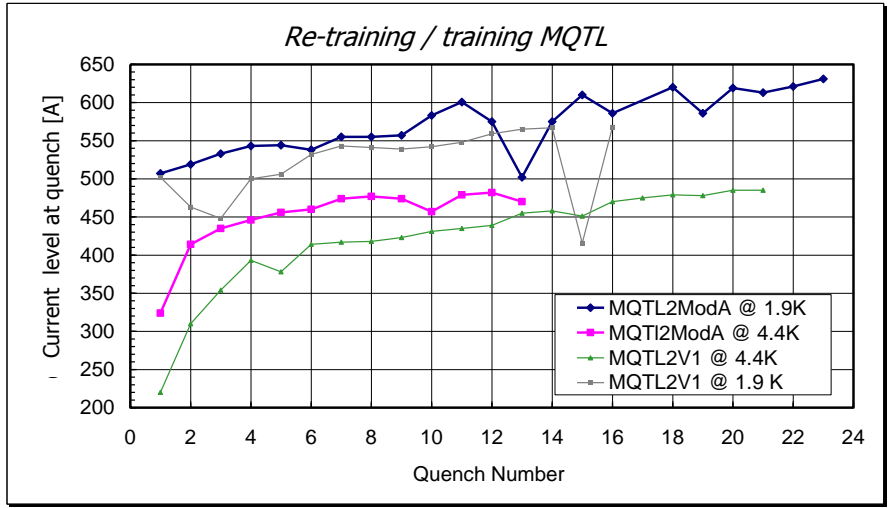

**Fig. 21:** Training tests of MQTL second prototype

An alternative method of providing the coils with the necessary pre-stress is based on iron blocks rather than scissor laminations. In the case of the superconducting combined magnet prototype developed for TESLA500 project, the iron was split into four sector blocks (see Fig. 22), whose radii were calculated to fit with the coil package and the shell at cold conditions [19]. The coils are glued together with glass-fibre spacers and wrapped around with a glass-fibre bandage. The main magnet is a quadrupole, and two corrector dipoles, horizontal and vertical, are glued around the quadrupole coils. All are cos-theta type windings.

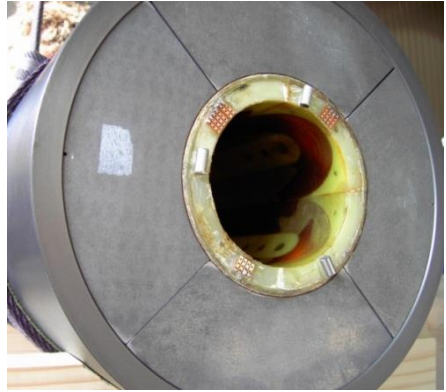

**Fig. 22:** Iron yoke split into four sector blocks

Figure 23 shows the hoop stress distribution when the coils are free, without any external support, compared with that when the coils are pre-compressed with an aluminium shell. In the former case, some tensile (positive) stresses appear in the region of contact of the coil and the central spacer, which is also the peak field region, i.e. the area more prone to trigger a quench. On the contrary, on fitting the shrinking cylinder, the full coil is under compression when the magnet is powered.

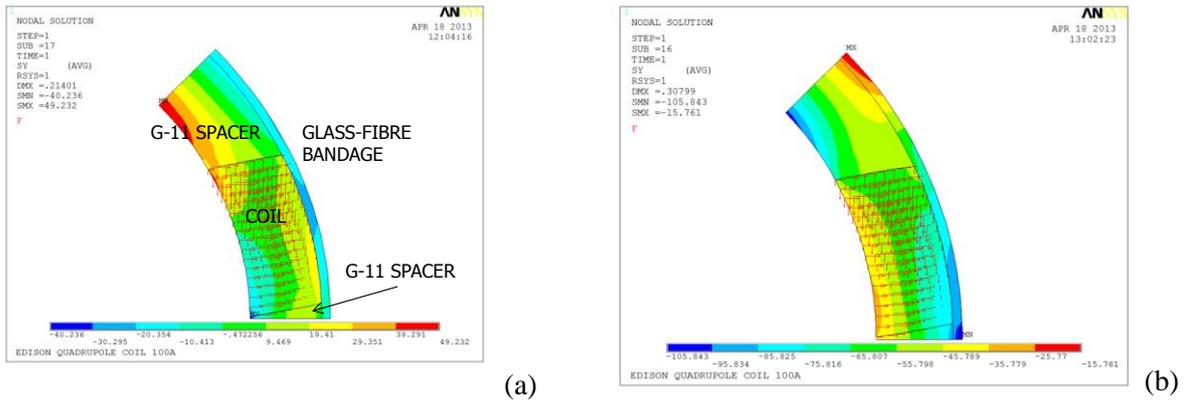

**Fig. 23:** Hoop stress distribution in the coil assembly of the TESLA500 magnet: (a) without pre-compression, (b) with pre-compression.

A very slow learning curve was recorded during the first training test (see Fig. 24), including a premature quench at about half the nominal current (100 A). Seventeen quenches were necessary to power the magnet at nominal current. The outer diameter of the shell was measured to evaluate the quality of the pre-compression, noting that two of the blocks had lost part of the pre-stress (see Fig. 25). The outer shells were disassembled and the interference was increased by gluing thin stainless steel sheets on the outer radius of the blocks with lower compression. It was checked that the shell outer diameter increased as expected, producing a symmetrical layout. Effectively, an important enhancement took place during the second training test: the third quench was already above nominal current. The magnet improved smoothly up to 130 A. In the re-training, the first quench was at a lower current, but still above nominal current. The most likely factors that still limit the magnet performance are the following:

- the absence of a structure to support the longitudinal electromagnetic forces, or
- the three layers of glued coils with glass-fibre spacers, which are relatively soft and have anisotropic properties. These are especially important when a bandage is wrapped around each layer of the finished coils: because it is applied manually, this could increase the inhomogeneity.

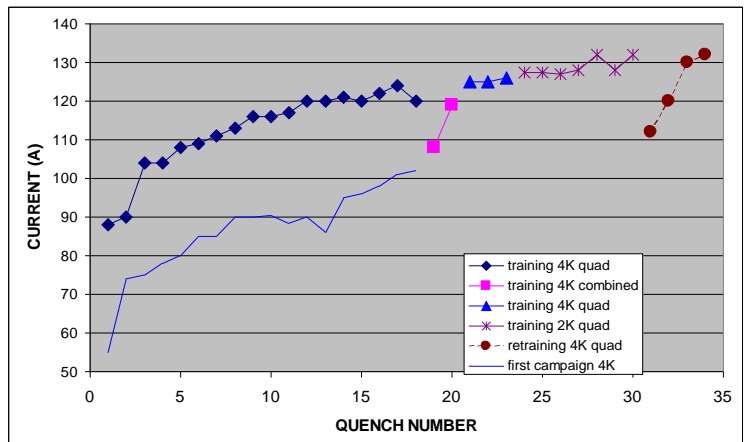

**Fig. 24:** Training tests of TESLA500 magnet prototype

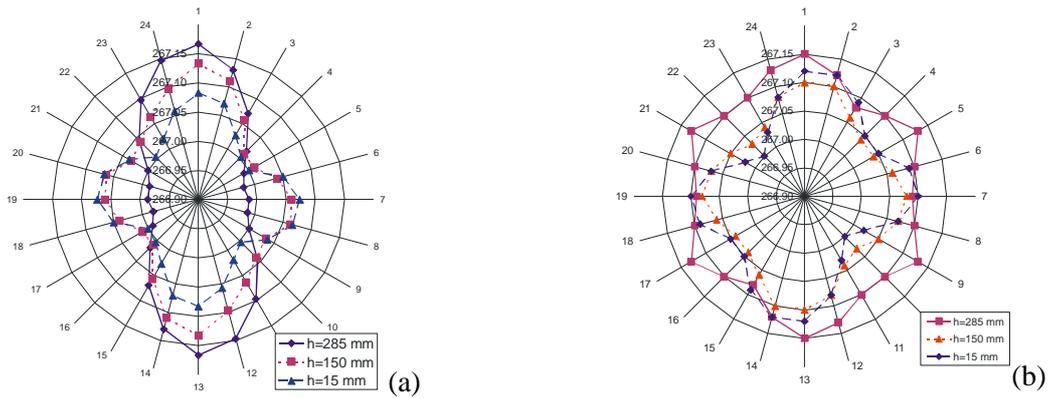

**Fig. 25:** Measurements of outer shell diameter at different angular positions before (a) and after (b) increasing the interference.

In short, coils of low field magnets are usually kept under compression by outer aluminium shells, fitted with some interference.

### 4.2.2 *High field magnets*

In this section, we refer to magnets as 'high field' as those with coil fields in the range from 4 to 10 T. Conductors are usually Nb−Ti cables with polyimide tape insulation, mostly of Rutherford type.

In the case of the Tevatron main dipole, the nominal field in the aperture is 4.4 T. When the coils are powered, the electromagnetic forces compress the cables azimuthally towards the mid-plane and radially against the external support structure (see Fig. 26). Assuming an infinitely rigid structure without pre-stress, the pole turn would move off about 100 $\mu$m, with a stress on the mid-plane of −45 MPa, at nominal current (see Fig. 27) [12].

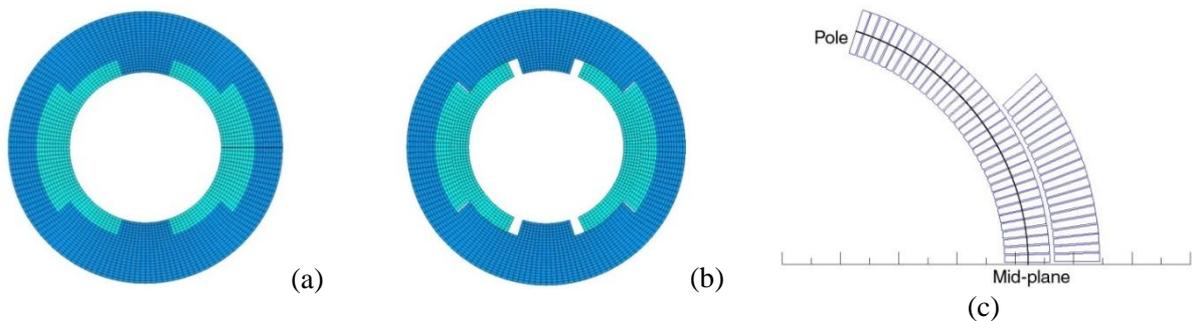

**Fig. 26:** (a) FEM model of Tevatron main dipole. (b) The coil moves off the pole when powered. (c) Coil cross-section: two layers of Rutherford cables [12].

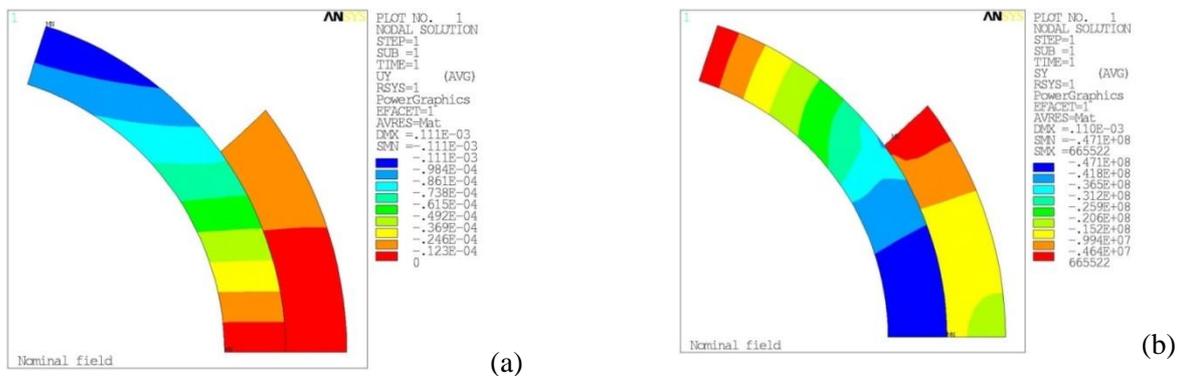

**Fig. 27:** Tevatron main dipole coil powered at nominal current: azimuthal displacements (a) and stresses (b) [12]

Figure 28 shows the azimuthal stress and displacement of the pole turn (i.e. the one with the highest field) in different pre-stress conditions at several current levels. The total displacement of the pole turn is proportional to the pre-stress. A full pre-stress condition (−33 MPa) minimizes the displacements and, probably, the quench triggering.

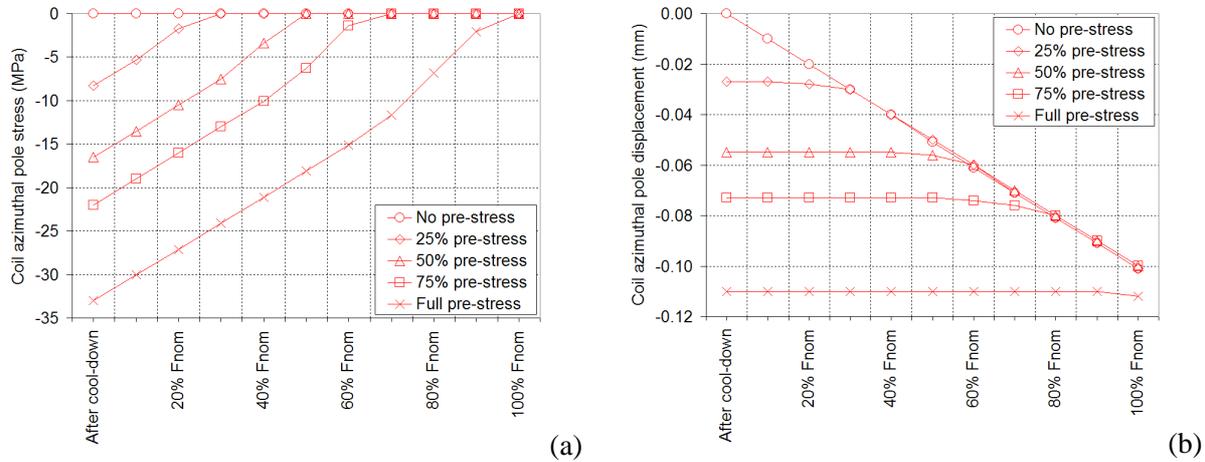

**Fig. 28:** Azimuthal stress (a) and displacement (b) of the pole turn of the Tevatron main dipole in different pre-stress conditions at several current levels [12].

The practice of pre-stressing the coil has been applied to all accelerator large dipole magnets: Tevatron [20], HERA [21], SSC [22, 23], RHIC [24] and LHC [25]. The pre-stress is chosen in such a way that the coil remains in contact with the pole at the nominal field, sometimes with a 'mechanical margin' of more than 20 MPa (see Fig. 29).

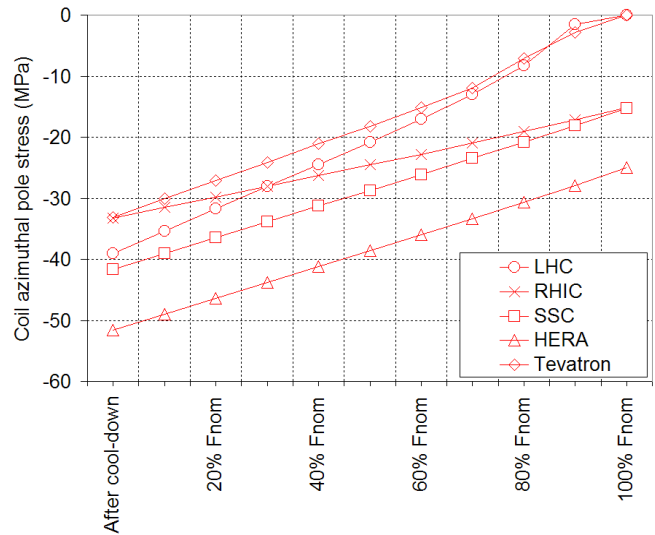

**Fig. 29:** Azimuthal stress at the pole turn for different coils of the main dipoles of large particle accelerators [12].

In high field magnets, the pre-stress is usually provided by means of collars. Collars were implemented for the first time in the Tevatron dipoles. Since then, they have been used in all but one (RHIC) of the high field cos-theta accelerator magnets and in most of the R&D magnets. They are composed of stainless steel or aluminium laminations of a few millimetre thickness. The collars take care of the Lorentz forces and provide a high accuracy for coil positioning. Shape tolerance is about ±20 $\mu$m. A good knowledge of the coil properties (initial dimensions and modulus of elasticity) is mandatory to predict the final coil status: both coils and collars deform under pressure.

Collars usually consist of two paired pieces with different geometries (see Fig. 30). The uncompressed coils are oversized with respect to the collar cavity dimension. The collars have holes or key slots which are aligned when both the collars and the coils are pressed at the nominal value. At that position, some bolts or keys are pushed through to lock the assembly. Once the collaring press is released, the collars experience a 'spring back' due to the clearance of the locking feature and deformation. The pre-stress may also change during cool-down due to the different thermal contraction of the collars and coils.

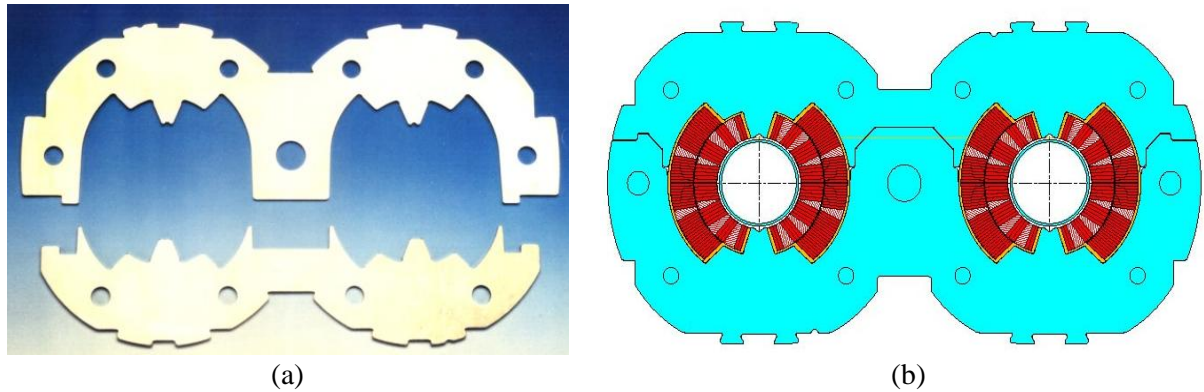

**Fig. 30:** Paired collars (a) and assembly with LHC main dipole coils (b) [26]

For fields above 6 T, it is usually necessary that the rest of the structure contributes to support the Lorentz forces. For instance, at nominal field, a LHC dipole experiences a horizontal force of 1.7 MN·m$^{-1}$ and a vertical one of −0.75 MN·m$^{-1}$ per quadrant. The stainless steel outer shell is split into two halves which are welded around the yoke at high tension (about 150 MPa) to withstand those forces. It is worth noting that when the yoke is placed around the collared coil, a gap (vertical or horizontal) remains between the two halves. This gap is due to the collar deformation induced by coil pre-stress. If necessary, during the welding process the welding press can impose the desired curvature on the cold mass. In the LHC dipole, the nominal sag is 9.14 mm.

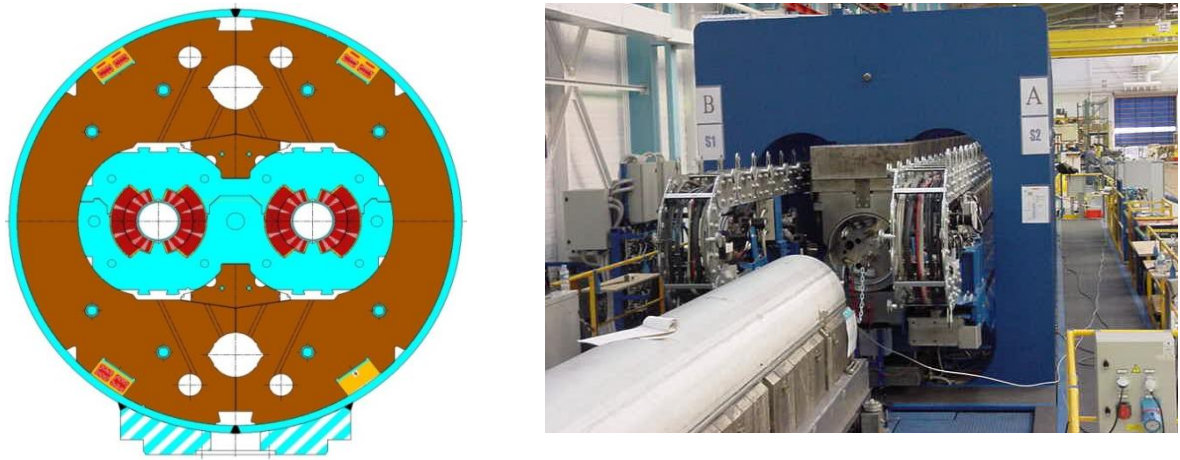

**Fig. 31:** (a) Cold mass of LHC main dipole. (b) Vertical press with automatic welding for the assembly of the outer shell [27].

End plates, which are applied after shell welding, provide axial support to the coil under the action of the longitudinal electromagnetic forces. A given torque may be applied to the end bolts. In some cases, the outer shell can also act as a liquid-helium container.

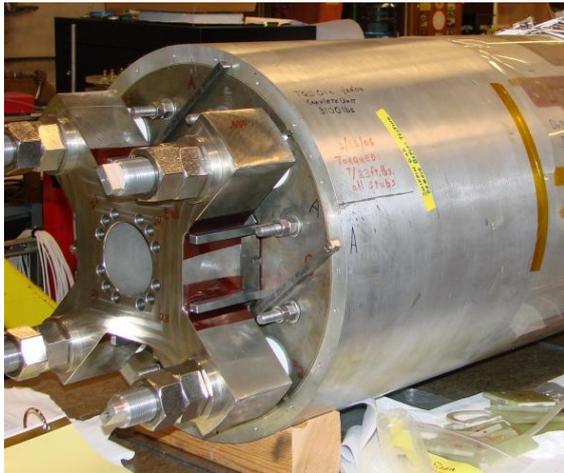 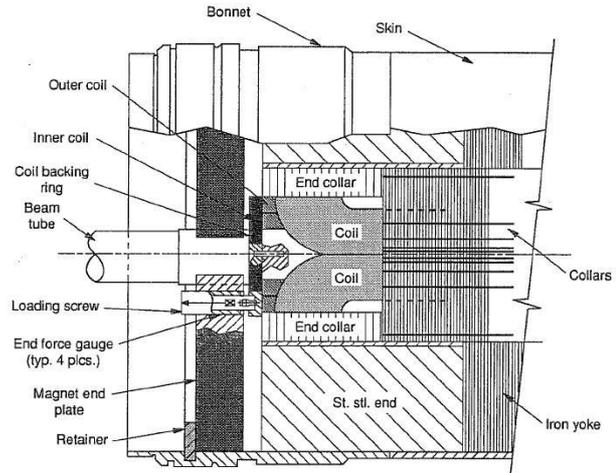

**Fig. 32:** (a) TQ quadrupole: thick rods hold the longitudinal Lorentz forces (courtesy of P. Ferracin). (b) Sketch of the SSC dipole end plates. (Courtesy of A. Devred.)

### 4.2.3 Very high field magnets

Let us consider as 'very high field' magnets those with coil peak fields beyond 10 T. These are R&D objects. All the collared magnets presented in the previous section are characterized by significant coil pre-stress losses (see Fig. 33):

− the coil reaches the maximum compression (about 100 MPa) during the collaring operation;
− after cool-down, the residual pre-stress is about 30–40 MPa.

What would happen if the 'required' coil pre-stress after cool-down were greater than 100 MPa? Following the same approach, the compression on the coil would be too high during the collaring.

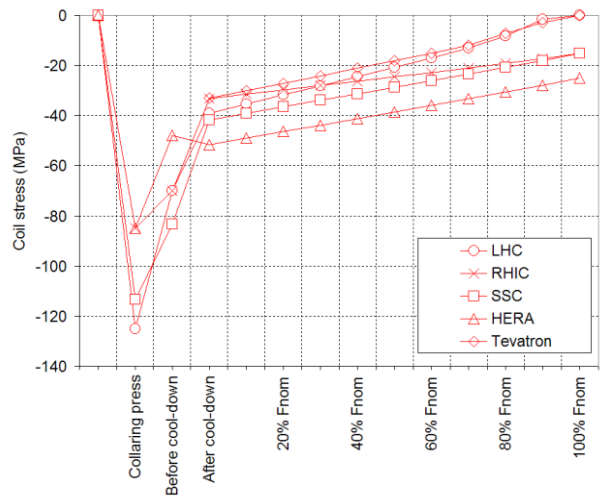

**Fig. 33:** Maximum compressive stress on the coils during the different assembly steps [12]

An alternative solution has been proposed and developed in the framework of the US LHC Accelerator Program (LARP). It is based on the use of bladders during the magnet assembly [28]. Figure 34(a) shows the TQ quadrupole cross-section. The coils are surrounded by the iron, which is split into four pads and four yokes, which remain open during all magnet operations. An outer aluminium shell contains the cold mass. The initial pre-compression is provided by water-pressurized bladders and locked by keys. During cool-down, the coil pre-stress significantly increases due to the high thermal contraction of the aluminium shell. Figure 34(b) shows how the maximum compressive stress on the coil is similar to that on the collared magnets, but this maximum takes place after cool-down and is available to counteract the electromagnetic forces. A small spring back occurs when bladder pressure is reduced, since some clearance is needed for key insertion.

R&D work is ongoing to prove that magnets assembled with this method:

− are able to provide accelerator field quality, and
− may be fabricated with lengths of several metres.

One of the magnets for the ongoing LHC upgrade is being designed following this approach: the MQXF quadrupole (140 T·m$^{-1}$ gradient in 150 mm aperture).

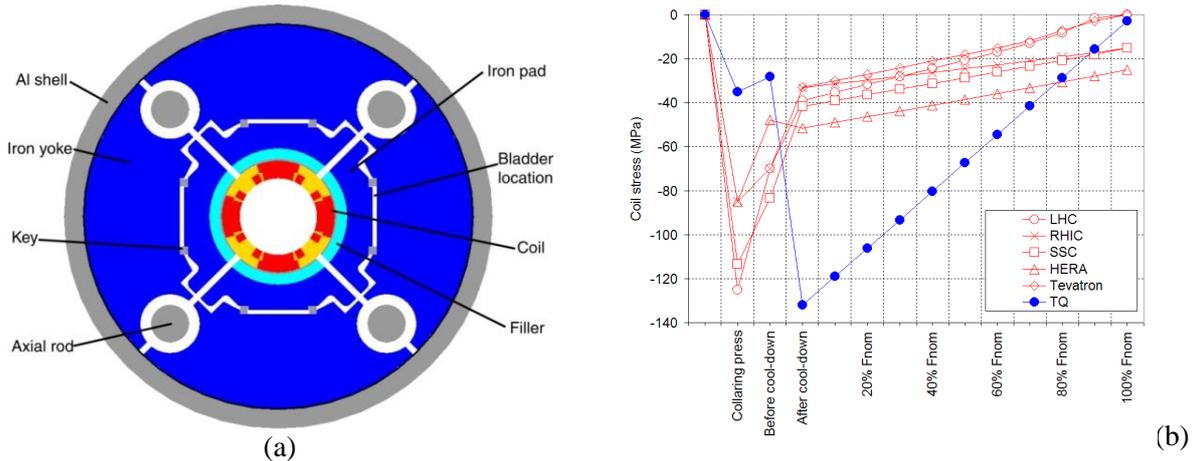

**Fig. 34:** (a) TQ quadrupole cross-section. (b) Maximum compressive stress on the coils during the assembly steps [12].

Another novel stress management system developed at Texas A&M University is based on intermediate coil supports [29]. Each coil block is isolated in its own compartment and supported separately (see Fig. 35). Lorentz forces exerted on multiple coil blocks do not accumulate, but rather are transmitted to the magnet frame by the Inconel ribs and plates. A laminar spring is used to pre-load each block.

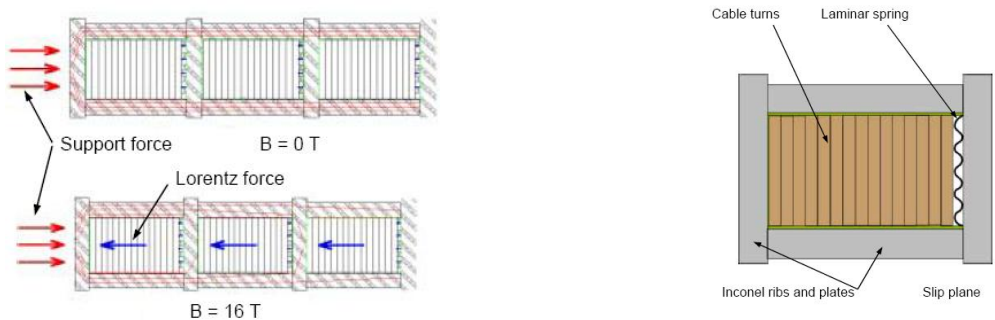

**Fig. 35:** (a) TAMU dipole coil blocks. (b) Detailed view of a coil block [30]

### 4.2.4 Pre-stress: controversy

As we have seen, the pre-stress aims to avoid the appearance of tensile stresses in the coils and limit the movement of the conductors. This raises the question: what is the correct value of the pre-stress?

In Tevatron dipoles, it was found that there was not a good correlation between small coil movements (<0.1 mm) and the magnet learning curve. In the LHC short dipole program, the coils were unloaded at 75% of the nominal current, without degradation in the performance. In LARP TQ quadrupoles, two different behaviours were detected (see Fig. 36):

- with low pre-stress, the coils were unloaded but kept a good quench performance;
- with high pre-stress, the learning curve was a stable plateau, but with a small degradation.

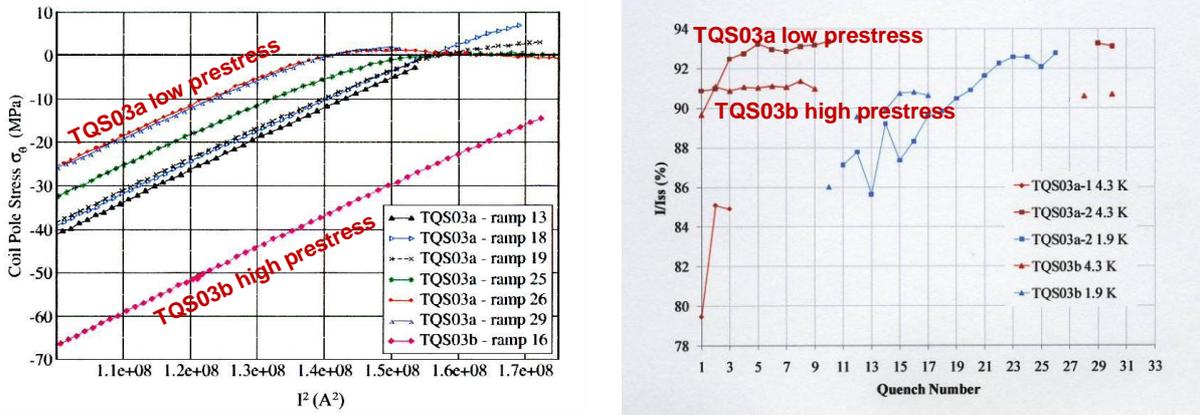

**Fig. 36:** (a) TQ coil stress vs. current. (b) TQ training tests with low and high pre-stress [12]

In LHC corrector sextupoles (MCS), a specific test program was run to find the optimum value of pre-stress [31]. Coils were individually powered under different pre-compressions immersed in the same field map as the magnet by means of a custom set-up (see Fig. 37). The conclusions were the following:

- The learning curve was poor in free conditions.
- Training was optimum with low pre-stress and around 30 MPa. However, degradation was observed for high pre-stress (above 40 MPa).
- The nominal pre-stress for series production was 30 MPa.

In conclusion, there is not an exact value for the correct value of pre-stress, but experience shows that too high a pre-stress can degrade the magnet performance.

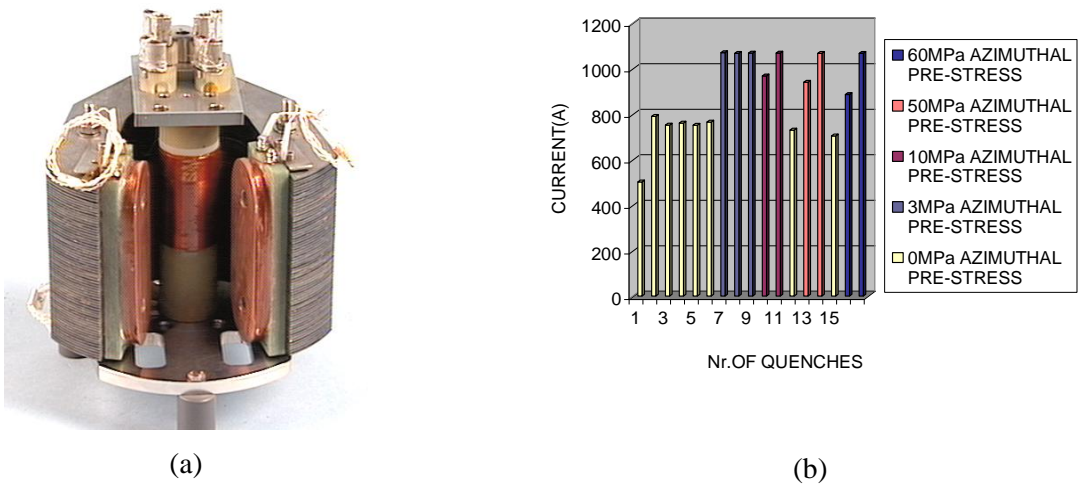

(a)                  (b)

**Fig. 37:** (a) Custom set-up to test individual MCS coils at different pre-stresses, provided by auxiliary superconducting coils. (b) Training tests with different pre-stresses [31].

## 5    Superferric accelerator magnets

The stress distribution in the coils of a superferric magnet is different from that in the cos-theta magnets: when powered, the coil experiences in-plane expansion forces, and it is usually attracted by the iron (see Fig. 38(a)). The force density is not as high as in cos-theta magnets, because fields are moderate.

In small magnets, simple support structures (such as wedges, see Fig. 38(b)) are sufficient to hold the coils and prevent wire movement, since coils are usually fully impregnated.

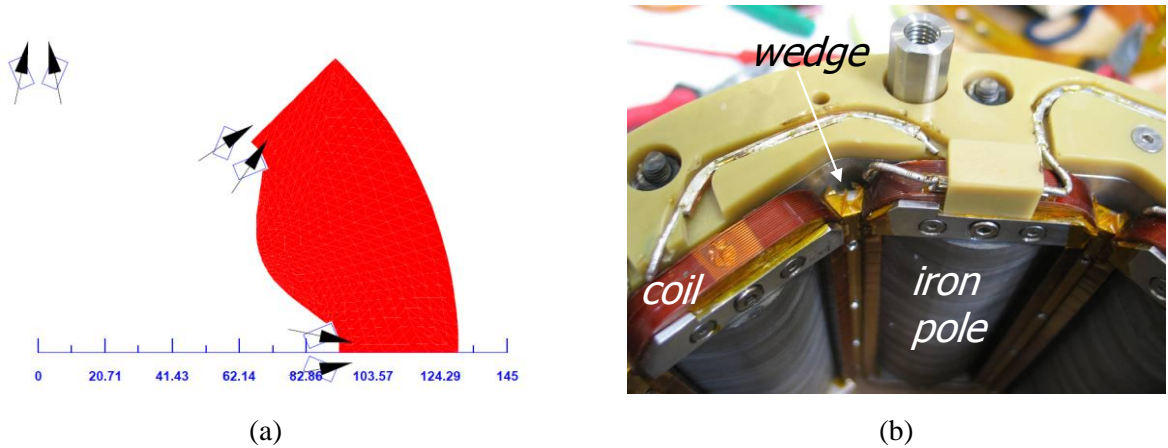

(a)　　　　　　　　　　　　　　　　　　(b)

**Fig. 38:** (a) Lorentz forces on superferric octupole coil blocks. (b) Support wedges in between two coils [32]

Large superferric magnets are very common in fragment separators (NSCL-MSU, RIKEN, FAIR) and particle detectors (SAMURAI, CBM). Usually, the iron is warm. Then, the Lorentz forces on the coil are counteracted by a stainless steel casing, which is also the helium vessel. In some cases, parts of these forces may be transferred to the external structure by means of low-heat-loss supports (see Fig. 39).

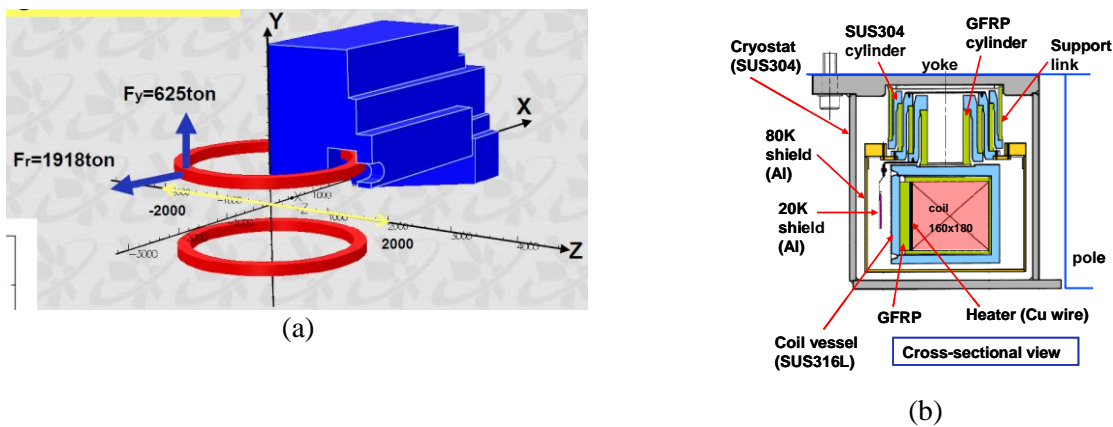

(a)　　　　　　　　　　　　　　　　　　(b)

**Fig. 39:** (a) Lorentz forces on Samurai magnet coils. (b) Cross-section of the cryostat [33]

## 6    Toroids

In toroids, as the magnetic pressure varies along the coil it is subjected to strong bending forces. If one wants to simplify the support structure, the following strategy must be followed [10]:

- Each coil experiences a net force towards the centre because the field is strongest there: it is wise to flatten the inner edge of the coil so that it leans on the support structure.
- The rest of the circumference will distort to a shape working under pure tension, where no bending forces are present. This tension must be constant around the coil. Assuming $R$ to be the distance to the centre and $\rho$ to be the local radius of the curvature of the coil, the condition for local equilibrium is given by

$$T = B(R)I\rho = \text{constant},$$
$$B(R) = B_0 \frac{R}{R_0}, \qquad (51)$$
$$\rho = \frac{\{1+(dR/dz)^2\}^{3/2}}{d^2R/dz^2} = \frac{TR}{B_0 R_0 I} = KR.$$

There is no analytical solution. Figure 40 shows a family of solutions. Toroids are used in large Tokamak fusion reactors, whose coil shapes resemble those depicted in Fig. 40. Nominal currents are usually very large. Indeed, the most commonly used cable is the so-called cable-in-conduit (CICC). The superconducting strands are free, enclosed within a stainless steel pipe, with a double objective: to host the coolant flow through the voids in between the strands, and to support the electromagnetic forces on the conductors. The use of this type of cable leads to some peculiarities regarding the mechanical calculations. We will review some of these aspects using the EDIPO magnet [34] as an example. It is a superferric dipole designed and fabricated to characterize cables for ITER coils. The nominal bore field is 12.5 T. The overall magnet length is 2.3 m. The Lorentz forces are huge: 1000, 500 and 400 tons in the horizontal, vertical, and longitudinal directions, respectively. The magnet is not collared. These forces are contained both by the low carbon iron laminations and by the outer stainless steel shell (see Fig. 41).

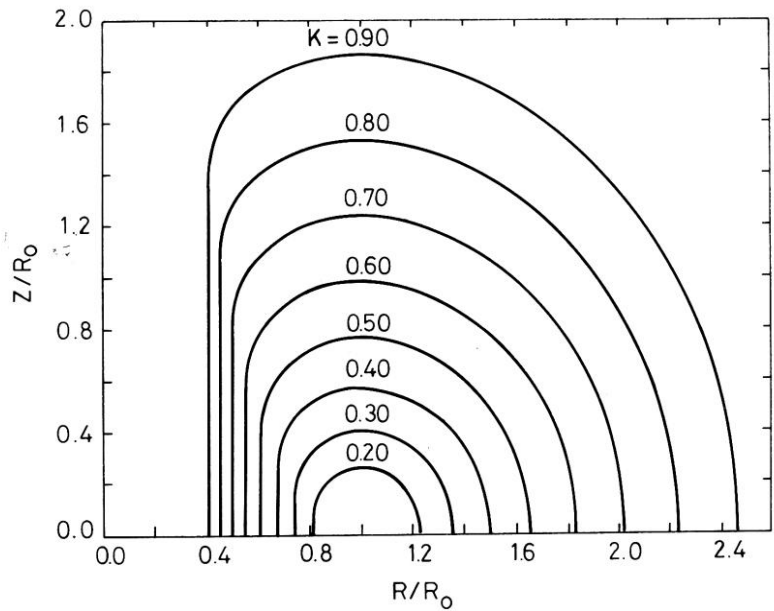

**Fig. 40:** Numerical solutions for toroid coil profile with constant tension and zero bending moment [10]

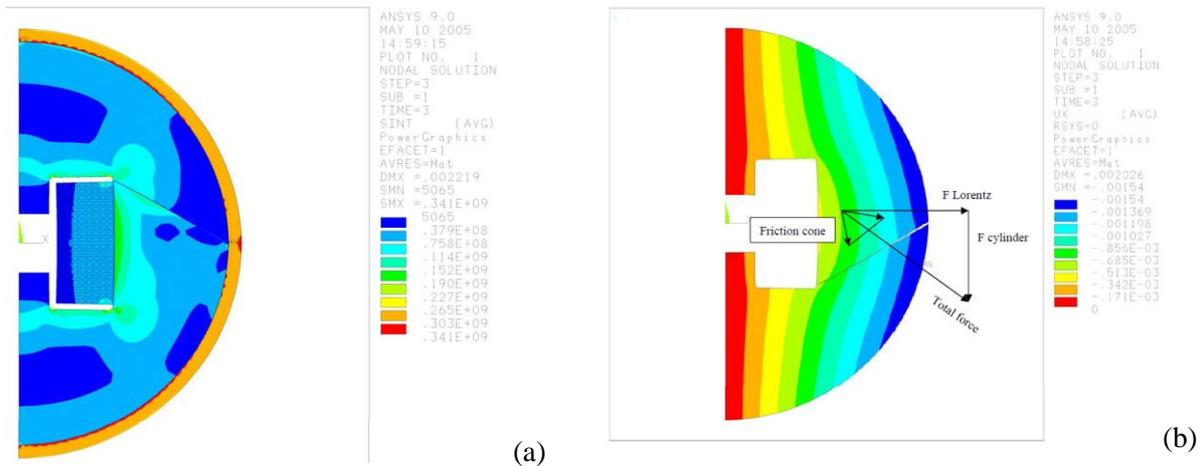

**Fig. 41:** (a) Tresca stress in EDIPO magnet. (b) Friction cone and horizontal displacements map. (Courtesy J. Lucas.)

It is worth pointing out that two different finite element models were used for the mechanical analysis:

− a general model, with few details, used to study the support structure deformation due to cool-down and Lorentz forces;

− a sub-model of the coil used to analyze the local stresses on the conductors, mainly in the insulation, which is wrapped around each CICC pipe. The Lorentz forces are transferred as internal pressures from the global model, and the contact with the support structure is modeled as a boundary condition (see Fig. 42).

This approach is very efficient, since there is no need to go into the details of the complete magnet model.

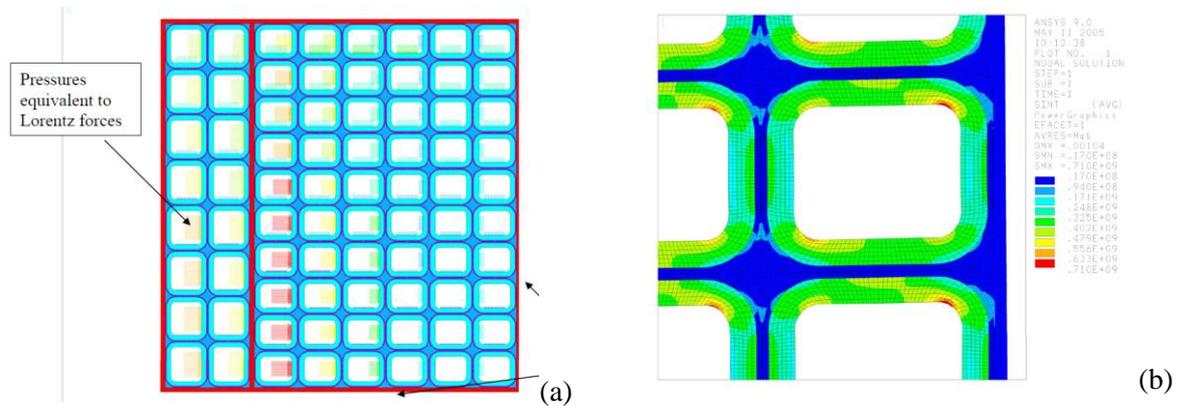

**Fig. 42:** (a) Sub-model used to analyze the stresses on the CICC insulation. (b) Tresca stress map. (Courtesy of J. Lucas.)

# 7 Case study: Advanced Molecular Imaging Techniques (AMIT) cyclotron

Let us finish our review on the mechanical design of superconducting magnets with a comprehensive example, the superconducting compact cyclotron of the AMIT project [35]. It includes a 4 T superconducting magnet with warm iron. The iron pole radius is 175 mm. The main components are depicted in Fig. 43. We will pay special attention to the mechanical design strategy and the flow of decisions.

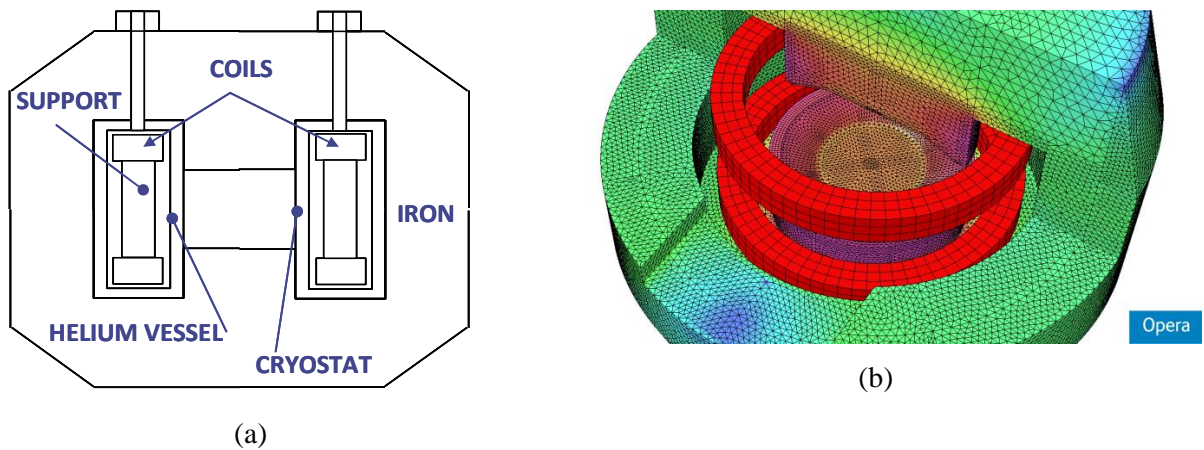

(a)

(b)

**Fig. 43:** (a) Cyclotron cross-section (courtesy of L. Garcia-Tabares). (b) 3D open model to show the coils and the iron poles. (Courtesy of D. Obradors.)

During the electromagnetic design, it was observed that the axial Lorentz forces between the coils could be attractive or repulsive, depending on their relative position and the operating current. It is easier to manage attractive forces, since the support structure would work under compression in that case, but it was not compatible with the dimensional constraints (compactness). Within the available space, the optimal design was that with the smallest repulsive forces, about 100 kN per coil.

In the conceptual design, the designer must analyze the requirements on the support structure (see Fig. 44).

- Radial Lorentz forces $F_r$ will induce a pressure on the winding. Since the coil is relatively thick, it is very likely that tensile (positive) radial stresses will appear in the inner layers of the coil. In any case, they will induce high hoop stresses in the superconductor. These forces will be held by an outer aluminium shell, fitted with a given interference. Due to the high thermal contraction of aluminium, the interference will be small and the assembly will be easy.

- Axial Lorentz forces $F_z$ will pull both coils towards the iron, thus inducing positive axial stresses in the windings. These forces will be held by a casing: when the coil is powered, it will press on the casing. Therefore, it is very important to guarantee the flatness of both contact surfaces, to avoid wire movements and, indeed, the quench triggering. These forces will induce bending moments at the corners of the support structure (holes are necessary to introduce the cyclotron vacuum chamber). A numerical model is needed to analyze the minimum fillet radius necessary to limit the stress concentration on those corners.

- The support structure will also be the helium vessel. The maximum pressure will take place in case of quench, when the helium suddenly boils off. A thermo-hydraulic model will be used to determine the necessary cross-section of the helium flow to limit the pressure and, indeed, the stresses induced on the vessel.

- Finally, one should choose the material. The structure will be made using non-magnetic stainless steel, which fulfils all the requirements: high magnetic field, low temperature operation, high stresses, and liquid-helium tightness. The best steel grades are 1.4429 and 1.4435, the second one being easier to be procured in small quantities on the market.

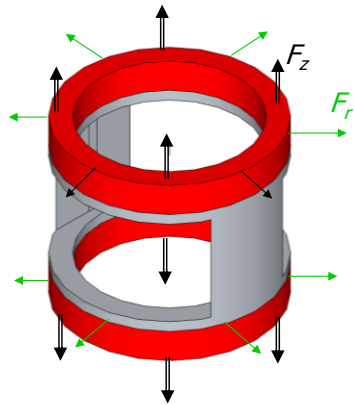

**Fig. 44:** Support structure for AMIT cyclotron: coils are inside the red casings, which are welded to the central part. (Courtesy of J. Munilla.)

Stresses in the coil and aluminium shell have been calculated in warm, cold, and energized conditions. Figure 45 shows the distribution of radial and hoop stresses. Radial stresses in the coil are always negative (compressive), whereas hoop stresses are positive, but are limited to 50 MPa in the coil and 150 MPa in the shell.

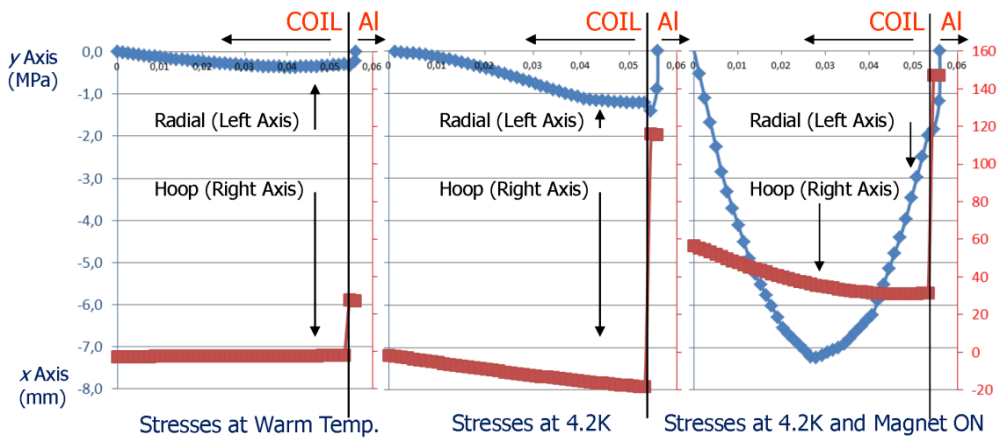

**Fig. 45:** Stresses on the coils at different assembly steps as a function of the distance to the coil inner radius. (Courtesy of J. Munilla.)

The stress distribution in the stainless steel support structure has been calculated using FEM. The maximum values are located in the corners (see Fig. 46) due to the bending moments induced by the axial electromagnetic forces. Their value is of the order of 80 MPa, which is an acceptable value for the steel.

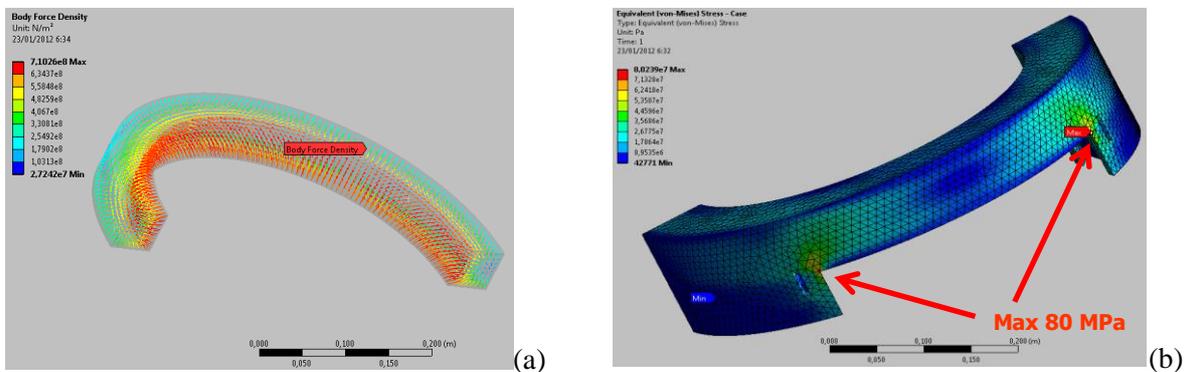

**Fig. 46:** (a) Lorentz forces on the coil. (b) Von Mises stresses in the coil casing (Courtesy of J. Munilla)

When the coils are not centred with the iron poles, some forces will arise. These forces have been calculated in three directions (see Fig. 47). The *x* and *y* horizontal axes are different due to the presence of the vacuum chamber hole through the iron yoke. The forces are in the direction of the misalignment with a positive slope, i.e. trying to increase the off-axis error.

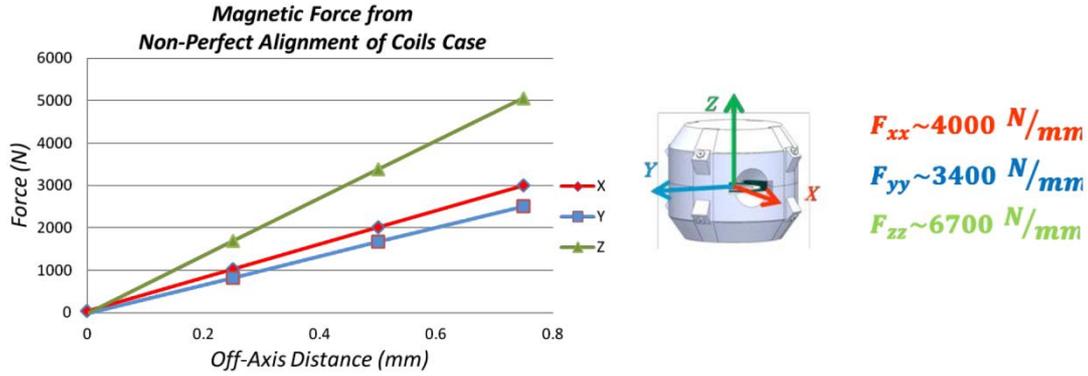

**Fig. 47:** (a) Magnetic forces vs. misalignments of coils with respect to the iron yoke (Courtesy of J. Munilla)

Stresses in the casing supports have been calculated in both centred and off-axis conditions (see Fig. 48). Upper rods will develop larger stresses because the coils hang from them.

In summary, the mechanical design interacts with the electromagnetic and the cooling calculations. The design decisions must take into account all these aspects to find the best trade-off. The first calculations are more general, even using analytical expressions at the very beginning. Once the concept is fixed, we come into the details, mainly with the use of numerical calculations and models. In the same way, a 2D approach is taken first, then a 3D design is realized, which is more time consuming.

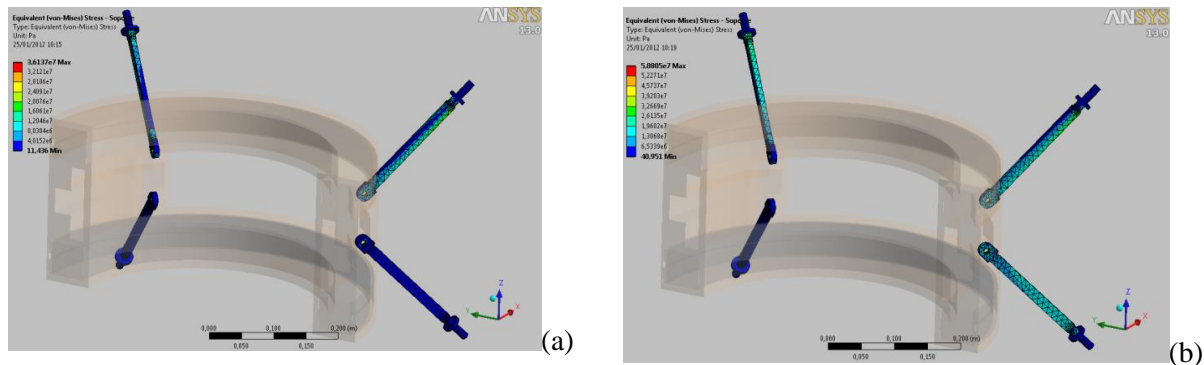

**Fig. 48:** (a) Von Mises stresses in the G10 rods at nominal current. (b) Von Mises stresses in the G10 rods at nominal current and a 0.5 mm misalignment in the *y*-direction. (Courtesy of J. Munilla.)

# 8 Measurement techniques

## 8.1 Stress

The most widespread technique for stress measurement is the *capacitive gauge*. The basic principle is to measure the variation of capacity induced by a pressure exerted on a capacitor. Figure 49(a) shows a typical layout of a capacitive gauge.

Let $S$ denote the area of the two parallel electrodes, with $\delta$ the thickness of the dielectric and $\varepsilon$ the electric permittivity; then the capacity $C$ is given by the following well-known expression:

$$C = \varepsilon S / \delta. \tag{52}$$

When a pressure σ is applied, the capacity will change as follows, due to the deformation of the dielectric:

$$C = \frac{\varepsilon S}{[\delta(1-\sigma/E)]}.\tag{53}$$

The gauges must be calibrated to achieve a good accuracy. The capacity can be measured as a function of pressure and temperature, as shown in Fig. 49(b) [36].

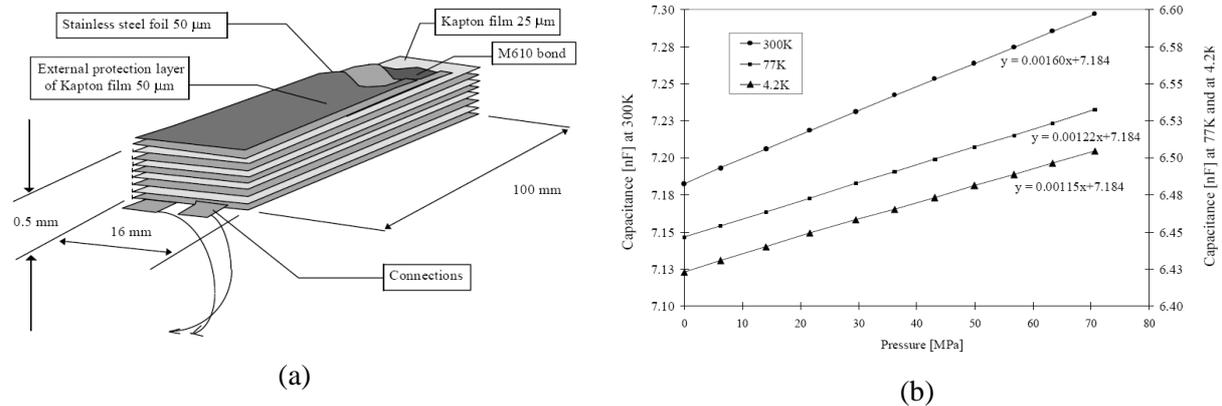

**Fig. 49:** (a) Typical layout of a capacitive gauge. (b) Capacitance measurement as a function of pressure at different temperatures [36].

### 8.2  Strain

The basic principle is to measure the variation of the resistance induced by a strain in a resistor [37]. The gauge consists of a wire arranged in a grid pattern bonded on the surface of the specimen (see Fig. 50). The strain experienced by the test specimen is transferred directly to the strain gauge. Once the gauges are glued, several thermal cycles may help to get rid of noisy or unstable measurements and to guarantee a good stress transfer. The gauge responds with a linear change in electrical resistance. The gauge sensitivity to strain is expressed by the gauge factor, which is the ratio of the resistance variation to the elongation:

$$GF = \frac{\Delta R/R}{\Delta l/l}.\tag{54}$$

The gauge factor is usually about 2. Gauges are calibrated by applying a known pressure to a stack of conductors or a beam. The temperature and magnetic field effects can be compensated for by measuring a nearby gauge which is not under stress.

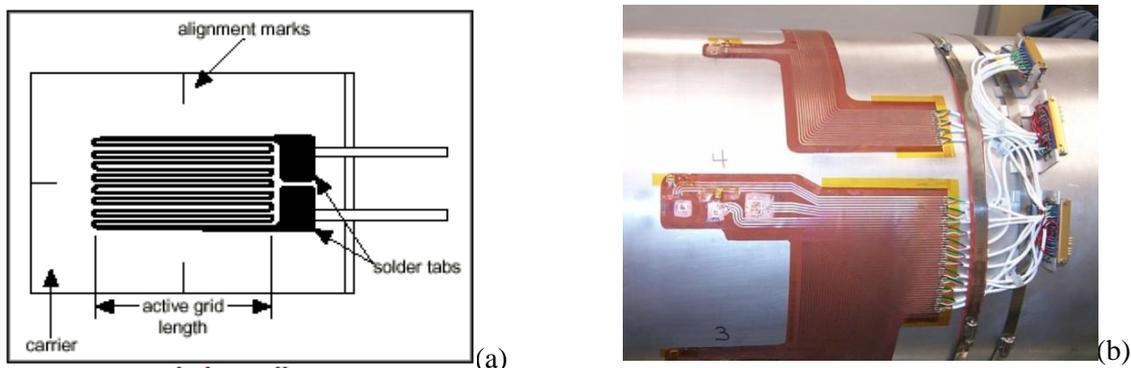

**Fig. 50:** (a) Typical layout of a strain transducer [37]. (b) Strain gauges glued onto an aluminium shell [12]

## 8.3 Coil physical properties

The elastic modulus $E$ is given by

$$E = \frac{d\sigma}{d\varepsilon} = \frac{d\sigma}{dl}l_0,  \quad (55)$$

where $\sigma$ is the applied stress, $\varepsilon$ is the specimen strain, $dl$ is the displacement, and $l_0$ is the initial length.

The elastic modulus $E$ is measured by compressing a stack of conductors, usually called a ten-stack, and measuring the induced deformation. The stress–displacement curve is not linear and presents a significant difference between the loading and unloading phases (see Fig. 51). The elastic modulus depends on the pressure applied and on the 'history' of the loading. It is also dependent on the temperature.

The thermal contraction is given by

$$\alpha = \frac{l_{w0} - l_{c0}}{l_{w0}}, \quad (55)$$

where $l_{w0}$ and $l_{c0}$ are the unloaded height of the specimen at room and cold temperature, respectively. Figure 51(b) shows a set-up to measure the thermal contraction by comparison with a well-known aluminium reference. It can be also evaluated using the stress loss in a fixed cavity.

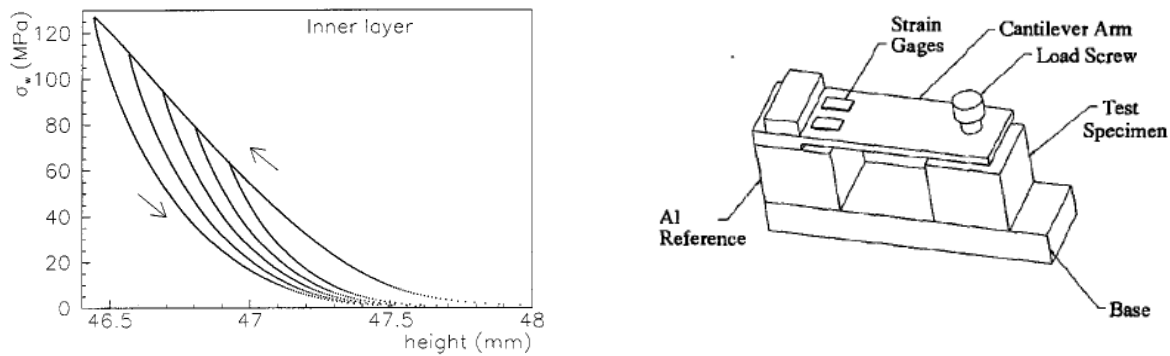

**Fig. 51:** (a) Typical stress–displacement curve of a cable stack [12]. (b) Set-up for thermal contraction measurement [38].

## Acknowledgement

The author warmly thanks Paolo Ferracin (CERN) for his support and helpful comments during the preparation of this lecture. His USPAS lectures [12] can be used by the interested reader to continue learning about the mechanical design of superconducting accelerator magnets.